\documentclass[prd,aps,preprint,amsmath,nofootinbib,amssymb,eqsecnum,showkeys,tightenlines]{revtex4-1}
\usepackage{slashed}
\usepackage{epsfig,latexsym,cancel,amssymb,amsmath,verbatim,mathrsfs}
\usepackage{color}
\usepackage{graphicx}

\def\ra{\rightarrow}
\def\L{\left(}
\def\R{\right)}
\def\wt{\widetilde}
\def\Ld{\Lambda}
\def\ld{\lambda}
\def\f{\frac}
\newcommand{\be}{\begin{equation}}
\newcommand{\ee}{\end{equation}}
\newcommand{\bea}{\begin{eqnarray}}
\newcommand{\eea}{\end{eqnarray}}
\newcommand{\ba}{\begin{array}}
\newcommand{\ea}{\end{array}}

\long\def\symbolfootnote[#1]#2{\begingroup%
\def\thefootnote{\fnsymbol{footnote}}\footnote[#1]{#2}\endgroup}

\newcommand{\beq}{\begin{equation}}
\newcommand{\eeq}{\end{equation}}

\begin{document}

\title{Scale-genesis by Dark Matter and Its Gravitational Wave Signal}

\author{Zhaofeng Kang}
\email[E-mail: ]{zhaofengkang@gmail.com}
\affiliation{School of physics, Huazhong University of Science and Technology, Wuhan 430074, China}

\author{Jiang Zhu}
\email[E-mail: ]{jackpotzhujiang@gmail.com}
\affiliation{School of physics, Huazhong University of Science and Technology, Wuhan 430074, China}

\date{\today}

\begin{abstract}

Classical scale invariance (CSI) may shed light on the weak scale origin, but the realistic CSI extension  to the standard model requires a bosonic trigger. We propose a scalar Dark Matter(DM) field $X$ as the trigger, establishing a strong connection between the successful radiative breaking of  CSI and DM phenomenologies. The latter forces the breaking scale to $\sim {\cal O}(\rm TeV)$. It brightens the test prospect of this scenario via gravitational wave, a sharp prediction of CSI phase transition (CSIPT), which is first order and has strong supercooling. Moreover, we carefully deal with some techniques which are commonly used to analyze CSIPT but maybe missed. In particular, we clarify the imprecision of Witten's formula used in the single field case to calculate the bubble nucleation rate and stress that the  essence of Witten's approximation is the validity of high temperature expansion.


\end{abstract}


\maketitle

\section{Introduction}

The origin and as well stabilization of the weak scale are the fundamental questions of the particle standard model (SM), and they lead the progresses of modern particle physics. Classical scale invariant (CSI) may shed light on these questions. The SM just contains two scales, the explicit negative mass parameter  $-\mu^2|H|^2$ in the Higgs potential, which accounts for the weak scale origin, and the dynamical scale $\Ld_{QCD}$ which determines the mass scale of the composite particles such as proton and neutron. Sending $\mu^2\ra0$ the SM becomes scale invariant at the classical level. However, CSI is violated at quantum level by anomaly, so hopefully a scale can be radiatively generated~\cite{Coleman:1973jx}. Moreover, 
CSI may be the symmetry that protects the weak scale free of the notorious fine-tuning, as long as the SM Higgs field does not couple to a heavy field with a sizeable strength, which generates a physical large quadratic correction to the Higgs field~\cite{CSI:natural}.

Nevertheless, the SM itself is not consistent with this symmetry, owing to the heaviness of top quark while the lightness of Higgs boson. To get a viable SM extension consistent with CSI, new elements probably bosons, should be introduced to trigger CSI spontaneously breaking (CSISB) at the quantum level. The trigger  is characterized by a relatively strong interaction with the scalon field, whose vacuum expected value (VEV) is the main order parameter of CSISB. It is tempting to conjecture that the new element is the missing piece of the SM, the dark matter (DM). Put it in another  way, a more inspiring way, DM spin is related to the weak scale origin. In such a framework DM plays a vital role, and it might ``explains" why DM should be there. In turn, due to the CSI, DM mass is no longer a dimensional parameter by hand; instead its mass origin is tied with CSISB. Moreover, as a trigger, the main interaction of DM is supposed to with the scalon field. Hence, this framework has its theoretical merits being a basis for DM model building~\footnote{Actually, studying DM in the CSI framework is not rare, and an incomplete list of references are collected in~\cite{Hur:2007uz,Guo:2014bha,Kang:2014cia,Guo:2015lxa,Ametani:2015jla,Ishida:2016fbp,
Hambye:2018qjv,Jung:2019dog,Guo:2018iix,Bian:2019szo,Karam:2015jta,Brdar:2018num}.}. We then consider a simple CSI extension to SM by two real singlet scalars, with one the scalon while the other one the trigger, the DM candidate at the same time. DM playing the role of the trigger is well expected from DM relic density, because the interactions of DM must be dominated by DM-scalon portal coupling so as to avoid the strong exclusion to the usual DM-Higgs portal coupling.

In the presence of multi scalars, one should carefully study the mechanisms of CSISB and electroweak (EW) spontaneously breaking. Different mechanisms are suitable for different parameter space of the dimensionless couplings; we refer to a nice review in Ref.~\cite{Chataignier:2018kay}. Confronting with the correct DM relic density and suppressed DM-nucleon recoil rate, we focus on two frequently studied cases:
\begin{description}
\item[ Gillard-Weinberg (GW) approach on the valley]  

The tree level potential admits a valley at some scale $\mu_{\rm GW}$ and one can apply the GW approach~\cite{Gildener:1976ih} to study CSISB. This approach treats the CSISB and the EWSB in a single step.

\item[ Coleman-Weinberg approach in the decoupling limit]  If the Higgs interactions with the scalon and trigger are irrelevantly small, CSISB by the scalon is reduced to the conventional single field case and one can apply the CW approach~\cite{Higgsportal}. Whereas the EWSB proceeds via the usual mechanism. 

\end{description}
Both scenarios leave viable parameter spaces, with scalon VEV at the multi-TeV scale and a heavy DM at the TeV scale.

DM triggers CSISB at zero temperature, and back to the early universe it triggers the CSI phase transition (CSIPT). It is known to be first order and tends to have a large supercooling, so abundant  gravitational wave (GW) may be produced during CSIPT. This has been studied by many groups~\cite{Witten:1980ez,Chiang:2017zbz,Marzo:2018nov,Prokopec:2018tnq,Ellis:2019oqb,Aoki:2019mlt,Mohamadnejad:2019vzg,Brdar:2019qut,Kubo:2016kpb}. Specified to our model, due to the requirements from DM phenomenologies, the CSISB scale is at the multi-TeV scale hence a CSIPT scale, by virtue of strong supercooling, from GeV to hundreds of GeV, which falls in the sensitivity region of space-based interferometer. Despite of the clear physical picture, there are some unclear technique points in analyzing CSIPT. As one of the focus, this article is devoted to clarifying these subtle aspects.

The paper is organized as follows: In Section II we study CSISB by virtue of the DM field and in the following section DM phenomenology is analyzed. In Section IV, V we detailedly study the CSI phase transition and the subsequent GW signals, respectively. Section VI contains the conclusions and discussions.

\section{Scale-genesis by scalar dark matter}

As stated in the introduction, in the CSI extension to the SM, a radiative CSISB trigger, a bosonic field providing the dominant quantum correction is indispensable. One simple option is a spin-1 gauge boson from a gauge group~\cite{CSI:b-l}, for instance, $U(1)_{B-L}$. It is tempting to consider that the missing part of the SM, the DM field $X$ does this job. We find that related scenarios are investigated before, in particular a spin-1 DM~\cite{VDM,Carone:2013wla,Karam:2016rsz,YaserAyazi:2019caf}, though the authors do not explicitly name the scenario. Ref.~\cite{Guo:2015lxa} for the first time studied the possibility of a spin-0 DM being the trigger. Anyway, DM as a trigger establishes a close connection between DM and CSISB, and their interplay gives rise to interesting phenomenologies. In this section, we will first set up the model and then detailedly analyze how the DM could trigger radiative CSISB.




\subsection{Simple (effective) model setup}

To build a realistic CSI model that could successfully accommodate the weak scale and as well the Higgs boson, it is better to consider that the weak scale is not the main scale for CSISB. Otherwise, the model will hit the Landau pole at a very low scale without an elaborate arrangement of the model~\cite{Dermisek:2013pta}. So, we will focus on the scenario that CSISB is dominated by a scalar field $S$. It is dubbed as the scalon field, and we will also term it scalon for short. As expected, the simplest extension with $S$ but without a trigger fails~\cite{Farzinnia:2013pga}. In this paper, we introduce a real singlet scalar DM as the trigger. The total Lagrangian of our model is
\begin{equation} \label{eq:aperp2.1} 
\begin{split}
\mathcal{L}_{tot}=\mathcal{L}_{\rm  SM}({\rm no~ Higgs~ potential})+K_{S,X}-V_{0}(H,S,X),
\end{split}
\end{equation}
where $K_{S,X}$ collects the kinetic terms for $S$ and $X$, and $V_{0}$ is the tree level potential,
\begin{align}\label{V0}
V_{0}=\ld|H|^4+\f{\ld_{hs}}{2}|H|^2S^2+\f{\ld_{hx}}{2}|H|^2X^2+\f{\ld_{sx}}{4}S^2X^2+\f{\ld_s}{4}S^4+\f{\ld_x}{4}X^4,
\end{align}
where $H$ is the SM Higgs doublet, and its neutral component  $H_0$ is decomposed as $H_0=\f{1}{{\sqrt{2}}}\L{{\rm Re}H_0+i{\rm Im}H_0}\R$. The successful radiative CSISB vitally relies on the scalon-trigger/DM coupling with a sizable $\ld_{sx}$. In order to stabilize $X$, we impose a $Z_2$ symmetry which only acts on the DM field: $X\ra -X$. Similar model has been studied in Ref.~\cite{Ghorbani:2017lyk,Brdar:2019qut}, but we will widely extend their discussions. Actually, we prefer to explain the above model as an effective model for the CSI framework having a DM trigger, and one can complete it in many different ways, e.g., the model based on local $U(1)_{B-L}$ considered in Ref.~\cite{Guo:2015lxa}.

Due to CSI, an accidental $Z_2'$ acting on $S$ emerges, $S\ra -S$. As a result, CSISB at the same time breaks $Z_2'$. On the other hand, it is well-known that the spontaneously breaking of a discrete symmetry would give rise to the domain wall problem. It is easily cured by the seesaw sector for neutrino masses,
\begin{align}
-{\cal L}_{N}=y_N \bar \ell H N+\f{\ld_{sn}}{2}SN^2+h.c.,
\end{align}
because it explicitly breaks $Z_2'$. As a side comment, in order to generate neutrino masses in the framework of CSI, a specific singlet scalon field $S$ may be always required, no matter via the normal seesaw mechanism or via radiative mechanism~\cite{Rnu,Karam:2015jta,Brdar:2018num}. To simplify the discussions, we assume that terms in the seesaw sector are irrelevant to CSISB, and it is true as long as $\ld_{sn}$ is sufficiently small.

\subsection{CSI radiative symmetry breaking}

Now we investigate CSISB in the model considered in the last subsection. The beautiful idea that generating a mass scale from a theory without an explicit mass scale is by S. Coleman and E. Weinberg in the celebrated paper in 1973~\cite{Coleman:1973jx}. Their basic observation is that CSI is anomalous and thus it is broken by quantum effects, hence generating a scale, equivalently driving the ground state of the system away from the origin. The first attempt is a scalar QED, a single scalar field (the scalon) charged under an Abelian gauge group with gauge coupling $g_X$. Then, given the hierarchy $1\gg g_X^4\sim \ld_s$ with $\ld_s$ the self-coupling of the scalon, the model succeeds in radiative CSISB via tree-loop balance in the perturbative region. However, the situation becomes more complicated if the scalon is a combination of several fields. Later E. Gildener and S. Weinberg proposed a method~\cite{Gildener:1976ih,GWscale} to reduce this complicated problem to the single field case, but it merely applies to the models whose couplings demonstrate a hierarchy such that the quantum effects only play a role near the flat direction of the potential. Since the two approaches work in different patterns of couplings, we will analyze radiative CSISB in the two scenarios separately. But before that it is more illustrative to start from the general way to deal with radiative CSISB in the general situations, which applies to cases with more than two fields.

\subsubsection{A general strategy for multi-field potential}

To study the radiative symmetry breaking of the model, we should start from the effective potential for the three classical scalar fields $\vec {\phi}_{\rm cl}=(h_{\rm cl},s_{\rm cl},X_{\rm cl})$ for ${\rm Re}H_0$, $S$ and $X$; for simplicity the subscript will be dropped. However, to our purpose the DM field is supposed to develop no VEV, which effectively reduces the 3-dimension field space to 2-dimension. Then, the tree level effective potential is directly read from the tree level potential Eq.~(\ref{V0}),
\begin{equation} \label{tree} 
\begin{split}
V^{(0)}(h,s)=\frac{1}{4}\lambda h^{4}+\frac{1}{4}\lambda_{s} s^{4}+\frac{1}{4}\lambda_{hs} h^{2}s^{2}.
\end{split}
\end{equation}
As the simplest two scalar system, this potential is shared by a variety of models, despite of different triggers. The tree level potential does not contain any explicit mass scale, so its minimum must be located at the origin.

To study if the quantum effects could realize CSISB, we incorporate the 1-loop corrections to $V^{(0)}$, which are encoded in the Coleman-Weinberg (CW) potential, 
\begin{equation} \label{CWpot} 
\begin{split}
V^{(1)}(h,s)=\frac{1}{64 \pi^2}\sum_{a}n_a m_a(h,s)^4\left[\log{\frac{m_a(h,s)^2}{\mu^2}}-C_a\right],
\end{split}
\end{equation}
where $\mu$ is the renomalization scale. Index $a$ runs over all particles coupling to the classical backgrounds, and the species $a$ contains internal degrees of freedom $n_a$, concretely 
\begin{equation} \label{eq:aperp2.8} 
\begin{split}
n_W=6,\ \ \ n_Z=3,\ \ \ n_h=1,\ \ \ n_{GSB}=3,\ \ \ n_t=-12,\ \ \ n_s=1,\ \ \ n_X=1.
\end{split}
\end{equation}
The subscripts successively denote the $W/Z$ boson, Higgs boson, Goldstone bosons (GSB), top quark in the SM, and the scalon, trigger/DM beyond the SM. $C_a$ are constants and in the $\overline{ \rm MS}$ scheme $C_a=\frac{5}{6}$  for a spin-1 field while  $C_a=\frac{3}{2}$ for a spin-0 or spin-1/2 field. Note that if the radiative symmetry breaking involves large couplings questionable in perturbativity, one should apply the renormalization group (RG) approach to improve the one-loop CW potential.

Let us turn to the key ingredients in the CW potential, the background field-dependent masses $m_a$. Most of them have simple analytical expressions,
\begin{equation} \label{eq:aperp2.11} 
\begin{split}
m^2_{GSB}=\lambda h^2+\frac{1}{2}\lambda_{hs} s^2,~~~\nonumber
m^2_X=\frac{\lambda_{hx}}{2} h^2+\frac{\lambda_{sx}}{2} s^2
\end{split}
\end{equation}
\begin{equation} \label{eq:aperp2.12} 
\begin{split}
m^2_{W}=\frac{g^2}{4} h^2,\ \ \ \ \ m^2_{Z}=\frac{g^2+g'^2}{4} h^2,\ \ \ \ \ 
m^2_t=\frac{y_t^2}{2} h^2,
\end{split}
\end{equation}
where $g$ and $g'$ are the gauge couplings for $SU(2)_L\times U(1)_Y$, and $y_t$ is the top quark Yukawa coupling. The three GSB modes in the SM cause the gauge dependent issue for the effective potential, but one may ignore it due to $m_{GSB}=0$ at least in the vacuum thus suppressed contributions to the CW potential. A consistent treatment may need a new gauge~\cite{Andreassen:2014eha,Andreassen:2014gha}; see a recent work employing this new gauge~\cite{Loebbert:2018xsd}. The SM physical Higgs boson mixes with the scalon, having mass squared matrix
\begin{equation} \label{sH:mass} 
\begin{split}
M^2_{h-s}= 
\left (
\begin{matrix}
  m^2_{hh}&m^2_{hs}\\
  m^2_{hs} &m^2_{ss}
  \end{matrix}
    \right )=
\left (
\begin{matrix}
  \frac{6\lambda h^2+\lambda_{hs} s^2}{2}&\lambda_{hs} h s\\
 \lambda_{hs} h s &\frac{\lambda_{hs} h^2+6\lambda_s s^2}{2}
  \end{matrix}
    \right ).
\end{split}
\end{equation}
The eigenvalues of  $M^2_{h-s}$ are given by
\begin{equation} \label{eq:aperp2.10} 
\begin{split}
m^2_{\phi_\pm}=\frac{1}{2}\left[{\rm Tr}M^2_{h-s}\pm\sqrt{\L{\rm Tr}M^2_{h-s}\R^2-4 {\rm Det}M^2_{h-s}}\right],
\end{split}
\end{equation}
and the lighter one $\phi_-$ is massless in the GW scenario.

In principle, to determine if the model realize radiative CSISB, one can use brute force to search the ground state of the general $V^{(0)}+V^{(1)}$ without special structure; the vacuum is labeled by the vacuum expectation value (VEV) of the classical backgrounds, $(v_h,v_s)$. But in practice this approach is useful only in the context of numerical studies~\cite{Chataignier:2018kay,Loebbert:2018xsd}. In some scenarios, discussed subsequently, the general potential can be reduced to the one-dimensional case where one can develop analytical understanding in the radiative CSISB.

Anyway, after finding the vacuum, the particle spectra of the Higgs-scalon system can be obtained from the mass squared matrix Eq.~(\ref{sH:mass}) with $(h,s)$ replaced by $(v_h,v_s)$. But it just gives the leading order result. This matrix receives substantial quantum corrections. In our studies due to the well separation between the CSISB scale and weak scale, the corrections mainly come from the trigger-DM coupling, so it is sufficient to just incorporate the loop correction to the scalon mass element. Quantum corrections may obviously modify the tree level masses and mixing. It has immediate implications to the physics that are sensitive to the mixing angle, for instance, DM-nucleon recoil. The spectra is characterized by the presence of another light CP-even Higgs boson different than the SM-like Higgs boson $h_{\rm SM}$, corresponding to the pseudo GSB (pGSB) of the spontaneously breaking of anomalous CSI. Nevertheless, it is not bound to be the lighter one, and in principle
 $h_{\rm SM}$ can be identified either with  the heavier one $\phi_+$ or the lighter one $\phi_-$.

\subsubsection{Higgs portal scenario by the Coleman-Winberg (CW) approach}

Let us first consider the scenario characterized by $0<-\ld_{hs}\ll1$. Then it is justified to switch the Higgs portal coupling $\ld_{sh}$ and study radiative CSISB in the dark sector along, which just contains the scalon $S$ and the trigger $X$ described by the following potential,
\begin{equation} \label{DS} 
\begin{split}
V_{DS}=\f{\ld_s}{4}S^4+\f{\ld_{sx}}{4}S^2X^2+\f{\ld_x}{4}X^4.
\end{split}
\end{equation}
$X$ should have no VEV, servicing as a spin-0 similarity to the gauge trigger in the original CW mechanism. One may ask why $S$ instead of $X$, which actually has similar couplings to $S$, is selected out as the scalon field, and in the Appendix~\ref{RGESOL} we give a short comment. Anyway, to study CSISB  the model is reduced to one single field case, and the tree level potential of the scalon field is $V^{(0)}_s(s)=\f{1}{4}\lambda_ss^4$.

Although not very necessary, we adopt the RG improved potential to incorporate quantum effects. This approach helps to determine the vacuum reliable in the perturbative region radiatively, and also to manifest  the difference between a spin-0 and a gauge trigger. In general, the RGE improved effective potential takes the form of~\cite{Quiros:1999jp}
\begin{equation}\label{}
V_s^{(1)}(s)=\f{1}{4}\lambda_s(t_s)G(t_s)s^4.
\end{equation}
The quantum effects encoded in the wave function factor $G(t_s)=\exp\left[\int_{0}^{t_s}dt'\gamma_s(t'))\right]$ merely gives a sub-leading contribution; $t_s=\log \f{s}{\mu_0}$ with $\mu_0$ the renormalization scale. As a matter of fact, here this contribution vanishes because of the coincidentally vanishing 1-loop anomalous dimension of the scalon field, $\gamma_s(t)=0$~\footnote{It is readily understood by nothing but that 1-loop correction to the two-point function of $s$ merely receives contributions from the bubble diagrams, a result of the $Z_2'$ symmetry.}. The major quantum effects are encoded in the running coupling $\ld_s(t_s)$, which is the solution of a set of coupled RGEs, whose beta functions are (in the $\rm MS$ scheme) given by
\begin{equation} \label{RGE} 
\begin{split}
&\beta_{\ld_s}=\f{9}{8\pi^2}\lambda_s^2+\f{1}{32\pi^2}\lambda_{sx}^2,
\\
&\beta_{\ld_{sx}}=\f{1}{4\pi^2}\lambda_{sx}^2+\f{3}{8\pi^2}\lambda_s\lambda_{sx}+\f{3}{8\pi^2}\lambda_x\lambda_{sx},
\\
&\beta_{\ld_x}=\f{9}{8\pi^2}\lambda_x^2+\f{1}{32\pi^2}\lambda_{sx}^2.
\end{split}
\end{equation}
Other contributions are suppressed by $\ld_{hs}^2\ll1$ in the Higgs portal limit.

The dark sector Eq.~(\ref{DS}) is different than the scalar QED where the trigger-scalaon coupling, namely the gauge coupling is the only parameter in the trigger sector, and the corresponding RGEs admit an analytical solution. But Eq.~(\ref{RGE}) involves not only trigger-scalon coupling $\ld_{sx}$ but also an additional coupling, the trigger/DM self-interaction $\ld_x$. Moreover, the $\beta$ function of the  trigger-scalon coupling receives a cross term $\ld_s\ld_{sx}$, which is absent in the scalar QED system. Consequently, Eq.~(\ref{RGE}) no longer has an analytic solution. In the Appendix~\ref{RGESOL} we approximate the above RGEs to the $\ld_s-\ld_{sx}$ system, thus admitting an analytical solution.

Fortunately, determination of the condition for radiative CSISB $d(V_s^{(1)}(s))/ds=0$ does not need to solve RGEs. Taking $\mu_0$ at the vacuum $v_s$, the extremum condition is translated to an equation between $\beta_{\ld_s}$, $\gamma_s$ if present and $\ld_s$ at $t_s=0$:  
\begin{equation}\label{}
\L \beta_{\ld_s}+4\ld_s(t_s)\R|_{t_s=0}=0 \Rightarrow   \ld_s(0)+\f{1}{128\pi^2}\ld_{sx}(0)^2+\f{9}{32\pi^2}\ld_{s}(0)^2=0.
\end{equation}
It is the phenomenon of dimensional transmutation: The scale $v_s$ is traded with a relation between dimensionless couplings. $\lambda_s(0)$ should be very small thus the $\ld_s(0)^2$ term negligible, leading to the relation between scalon self-coupling and trigger-scalon coupling at $t_s=0$,
\begin{equation}\label{}
\lambda_s(0)\approx-\f{1}{128\pi^2}\ld_{sx}(0)^2.
\end{equation}
It reproduces the well-known tree-loop hierarchy, but it is $\ld_{tree}\sim \ld_{loop}^2$ rather than $\ld_{loop}^4$ in a scalar QED theory. Turning on the Higgs portal coupling leads to EWSB, which does not proceed radiatively, but as in the usual SM via a negative mass parameter $\mu_h^2=-\lambda_{hs}v_s^2/2$. It is an accidental result of the negative Higgs portal coupling between the scalon and SM Higgs doublet. The resulting weak scale is expressed as
\begin{equation}
v_h=\sqrt{\f{\mu_h^2}{\lambda}}=\sqrt{\f{-\lambda_{hs}}{2\lambda}}v_s,
\end{equation}
fixed to be $v_h=246$ GeV.

After pining down the vacuum, now we present the spectra. The heavy trigger or DM mass squared is $m_X^2\approx \f{1}{2}\lambda_{sx}v_s^2$. In the strict Higgs portal limit, there is no mixing between the SM Higgs boson and scalon. Then the former gets mass as usual in the SM, while the latter as a pGSB gains a mass purely from quantum effects,
\begin{equation}\label{}
m_s^2=\f{d^2}{ds^2}[V_s^{(1)}(s)]_{s=v_s}=b_Xv_s^2,
\end{equation}
with $b_X\approx\f{1}{32\pi^2}\lambda_{sx}(0)^2$, the main part of $\beta_{\ld_s}$. But the Higgs portal coupling generates a small mixing between them, and we should calculate the spectra from the mass squared matrix Eq.~(\ref{sH:mass}), with $m_{ss}^2\ra m_{ss}^2+b_Xv_s^2$ capturing the dominant quantum effect. Usually $m_s$ is a good approximation to the actual scalon mass even after taking into the small mixing effect. 


\subsubsection{GW region by GW approach}

Let us move to the scenario characterized by $V^{(0)}(\phi_i)\gg V^{(1)}(\phi_i)$, and then quantum corrections numerically do not matter except for the places in the field space where $V^{(0)}\approx 0$. In the CSI models, the valley or flat direction is such kind of place. It is determined by the nontrivial solution to the extremum equation $dV^{(0)}/d\phi_i|_{\mu_{GW}}=0$, which, specific to our model, are given by
\begin{equation} \label{GW} 
\ld h^2+\f{1}{2}\ld_{hs}s^2=0,\quad \ld_s s^2+\f{1}{2}\ld_{hs}h^2=0. 
\end{equation}
Their solution is denoted as $\phi \vec{N}$, corresponding to lines of degenerate local minimum (thus flat), which points to a definite direction $ \vec{N}$ in the filed space, but leaving the modulus $\phi$ free. Eq.~(\ref{GW}) admit nonzero solutions only for a special relation (GW relation) among the dimensionless couplings,
\begin{equation} \label{flat} 
\f{h^2}{s^2}=-\f{1}{2}\f{\ld_{hs}}{\ld}=-2\f{\ld_s}{\ld_{hs}}\Rightarrow \ld_{hs}=-2\sqrt{ \ld\ld_s}<0. 
\end{equation}
The GW relation can be regarded as a renormalization condition at the GW scale $\mu_{GW}$~\footnote{Gildener proved that GW relation is feasible via choosing $\mu_{GW}$ in the RGE~\cite{GWscale}: Starting from a potential without a valley at a generic scale, it can flow to the one with a valley at  $\mu_{GW}$. Hence the GW relation does not require fine tuning of couplings. However, this procedure works only for the couplings not far off the GW relation, otherwise RGEs fail in driving the couplings to satisfy the GW relation at some scale. }. Now the flat direction is expressed as the following, 
\begin{equation} \label{eq:aperp2.14} 
\begin{split}
\left (
\begin{matrix}
  h  \\
 s
  \end{matrix}
    \right )
    =
   \phi \vec{N}=
   \phi\left (
\begin{matrix}
  \cos\alpha
  \\
 \sin\alpha
  \end{matrix}
    \right )=
      \frac{\phi}{\sqrt{\lambda^{\frac{1}{2}}+\lambda_s^{\frac{1}{2}}}}\left (
\begin{matrix}
  \lambda_s^ {\frac{1}{4}}
  \\
 \lambda^ {\frac{1}{4}}
  \end{matrix}
    \right).
\end{split}
\end{equation}
In particular, we are interested in small mixing $\alpha\ra \pi/2$ and therefore Eq.~(\ref{flat}) implies the hierarchy
\begin{equation} \label{hierarchy} 
0<\ld_s\ll-\ld_{hs}\ll \ld.
\end{equation}
Like in the Higgs portal scenario, again we need a very small and moreover negative $\ld_{hs}$, but the underlying reasons are not the same.

The Radiative correction will lift the flat direction and create a local minimum at some $\phi$. To see this we rewrite the tree-level effective potential Eq.~(\ref{tree}) plus the radiative correction along the flat direction in terms of $\phi$, i.e., reducing the potential to the one dimensional case, 
\begin{equation} \label{eq:aperp2.15} 
\begin{split}
 V^{(0)}+ V^{(1)}(\phi\vec{N})=A\phi^4+B\phi^4 \log\frac{\phi^2}{\mu_{GW}^2},
\end{split}
\end{equation}
where $A$ and $B$ are dimensionless loop functions defined as
\begin{equation} \label{eq:aperp2.16} 
A=\frac{1}{64\pi^2}\sum_a n_a m_a(\vec N)^4[ \log m_a(\vec N)^2-C_a],
~~~~~ B=\frac{1}{64\pi^2}\sum_a n_a m_a(\vec N)^4,
\end{equation}
which only  depend on the tree level couplings. As expected, the tree level potential vanish along the flat direction. By finding the extremum of  $V^{(0)}+V^{(1)}(\phi\vec{N})$ we know that the modulus is fixed to be at the position related to the GW scale as,
\begin{equation} \label{eq:aperp2.17} 
\begin{split}
\log\frac{\left\langle\phi\right\rangle^2}{\mu_{GW}^2}=-\frac{1}{2}-\frac{A}{B}.
\end{split}
\end{equation}
The weak scale and the scalon scale respectively are given by $v_h=\left\langle\phi\right\rangle\cos\alpha=246~\rm GeV$ and  $v_s=\left\langle\phi\right\rangle\sin\alpha$.

Using the GW relation, it is ready to show that the Higgs-scalon mass squared matrix Eq.~(\ref{sH:mass}) presents the GSB (or the consequence of flat direction), $m_h{\phi_-}=0$. And quantum corrections add a new piece $8B\langle\phi\rangle^2\sin^2\alpha$ to the $m_{ss}^2$ element. In the small mixing limit $\alpha\ra \pi/2$, it is just the mass squared of the pGSB (as a reminder, pGSB is not necessary $\phi_-$). Then, to guarantee that the extremum is a minimum, $B>0$ is required and it means
\begin{equation} \label{eq:aper2.20} 
\begin{split}
6m_W^4+3m_Z^4+m_h^4+m_X^4-12m_t^4>0,
\end{split}
\end{equation}
which yields the lower bound on DM mass, $m_X>316.48 {\rm GeV}$. Since $\alpha\ra\pi/2$, the main component of pGSB comes from the singlet. While the SM Higgs boson takes up the dominant fraction of the SM-like Higgs boson $h_{\rm SM}$, whose mass squared neglecting a small shift from loop is well approximated by
\begin{equation} \label{eq:aperp2.19} 
\begin{split}
m_{h_{\rm SM}}^2\approx {\rm  Tr}(M_{h-s}^2)=(3\lambda+\frac{\lambda_{hs}}{2}) v_h^2+(3\lambda_s+\frac{\lambda_{hs}}{2}) v_{s}^2= \L2\ld-\ld_{hs}\R v_h^2.
\end{split}
\end{equation}
Since $\ld_{hs}\ll \ld$, the above expression basically is the same with the one predicted in the SM. The Higgs-scalon mixing angle $\theta$ coincides with $\alpha$ at tree level, but it may be subjected to strong  radiative correction, in particular when $8B\langle\phi\rangle^2$ is close to $m_{h_{\rm SM}}^2$. 
%

\section{The scalar dark matter via the scalon portal}

In this section we focus on the other face of the trigger, the DM candidate. It mainly interacts with the scalon field (hence scalon portal), which is supposed to determine the DM relic density. Its  correct value $\Omega h^2\simeq 0.12$ limits the scale of CSISB. Moreover, the scalon-SM Higgs boson mixing is subjected to constraints from the null DM direct detention results. In a word, the trigger being a DM has strong impacts on radiative CSISB.

First of all, we collect the relevant terms for the DM dynamics. We introduce a more illustrative notation to label the eigenstates of $M^2_{h-s}$. Let $O(\theta)$ be the orthogonal matrix that diagonalizes the full $M^2_{h-s}$, and its mass eigenstates $\phi_\pm$ are renamed as $H_i=(h_{\rm SM}, {\cal S})$, related to the flavor states via
\begin{align}\label{massbasis}
h_{\rm SM}= \cos\theta h+\sin\theta s,\quad {\cal S}= \cos\theta s-\sin\theta h.
\end{align}
Note that $h_{\rm SM}$ can be identified with $\phi_+$ or $\phi_-$, depending on the relative size of diagonal elements of $M^2_{h-s}$ at loop level. Then, the relevant interactions in the mass basis are collected in the following Lagrangian
\begin{align}
-{\cal L}_{X}=\f{1}{4}X^2{\cal S}^2+ \f{A_{i}}{2} X^2H_{i}+\f{\ld_{ij}}4 X^2H_{i}H_j+\f{y_{iq}}{\sqrt{2}}H_i\bar q q.
\end{align}
where $A_i=\ld_{hx}v_h O_{1i}+\ld_{sx}v_s O_{2i}$ and $y_{iq}=y_q O_{1i}$, with $O$ specified in Eq.~(\ref{massbasis}).


\subsection{ DM relic density via freeze-out: CSIPT scale not far above TeV}

If $X$ is the unique DM component, we have to guarantee that its relic density is correctly produced. Assuming an ordinary thermal history for $X$, its relic density is determined by the usual freeze-out dynamics, which requires that $X$ should have an annihilation cross section times the relative velocity at the freeze-out epoch $\langle \sigma v\rangle\simeq 1$ pb.

The complete list of DM annihilation channels is long, for instance, into the various SM species via the Higgs-portal and as well into a pair of scalon. But it is well known that DM direct detection, discussed in the following subsection, compels us to consider a very small coupling between $X$ and the SM Higgs doublet. Subdominant of the SM Higgs portal greatly simplifies the dynamics of DM at the early universe, and the dominant annihilation channel of DM is $XX\ra {\cal S}{\cal S}$. It has a cross section times velocity~\footnote{As noticed in Ref.~\cite{Kang:2014cia,Guo:2015lxa}, for a fermionic DM, the common mass and annihilation dynamics from a single interaction can make the above estimation independent on $\ld_{sx}$.} 
\begin{equation}\label{relic}
\langle\sigma_{XX}v\rangle\simeq\f{ \ld_{sx}^2}{64\pi}\f{1}{m_X^2}=0.89 {\rm pb}\times\L
\f{\ld_{sx}}{2.0}\R \L\f{3{\rm TeV}}{v_s}\R^2
\end{equation}
where we have taken the limit $m_X\gg m_{\cal S}$. It is seen that the scale of $v_s$ cannot be very high confronting the perturbative bound on $\ld_{sx}<\pi$, which in turn means the heaviest DM $\simeq 4.4$ TeV capable of having a correct relic density.

A long comment deserves attention. The ordinary freeze-out dynamics of DM may be violated by supercooling CSIPT (discussed in the later section). Before the CSIPT, all particles including DM are massless, so DM is tightly coupled to the plasma. DM just gains mass after CSIPT. But if it is strongly supercooled, it is possible that the PT completion temperature $T_*<T_f\sim m_X/20$ with $T_f$ the estimation on the decoupling temperature of normal DM. This means that DM number density is not frozen at $T_f$ but at $T_*$, when DM gains a heavy mass much above the plasma temperature $\sim T_*$, thus failing to enter the new plasma inside the bubble, which expands to occupy the space dwelling in the metalstable vacuum. However, if the freeze-out dynamics indeed fails is a complicated question. One reason is that the reheating after CSIPT probably will heat the universe to a very high temperature and therefore DM may be thermalized again. A solid discussion is beyond the scope of this work, and we leave this very interesting topic to a specific publication. Here we just assume that the ordinary freeze-out still works.




%
%
%
%
%
%

\subsection{DM direct detection bounds}

The DM-nucleon elastic scattering is mediated by the Higgs bosons $H_i$, and they can be integrated out, generating the effective operators between DM and quarks, $a_{q}X^2\bar q q$ with
\begin{align}\label{aq}
a_{q}=\f{y_q}{\sqrt{2}}\f{A_i}{2}O_{1i}\f{1}{m_{H_i}^2}=m_q\sum_i\L \f{\ld_{hx}O^2_{1i}}{2m_{H_i}^2}
+\f{v_s}{v_h}\f{\ld_{sx}O_{1i}O_{2i}}{2m_{H_i}^2}\R,
\end{align} 
where the first and second are the contributions from the Higgs portal and scalon portal, respectively.
Then, the DM-nucleon elastic scattering cross section is $\sigma_{\rm SI}=\f{4}{\pi}\mu_p^2 f_p^2$~\cite{SDM}, with 
\begin{equation}\label{sigmaSI}
f_p=\f{m_p}{2m_X}\sum_q\f{a_q}{m_q}f^{(p)}_{T_{q}}, 
\end{equation}
where $\mu_p\approx m_p$ the reduced mass and $\Delta^p =\sum_{q=u,d,s} f_{T_q}^{(p)}+\sum_{q=t,b,d} f_{T_q}^{(p)}\approx 0.35$ encoding the nuclear factors. The direct detection upper bounds strongly limit the size of $A_i O_{1i}/m_{H_i}^2$. In particular, the usual SM Higgs portal coupling $\ld_{hx}$ must be highly suppressed. So, we can just retain the scalon-portal contribution in $f_p$, to derive  
\begin{align}\label{scalon}
f_p \approx \ld_{sx}\f{m_p}{8m_X}\f{v_s}{v_h}\sin2\theta\L\f{1}{m_{h_{\rm SM}}^2}-\f{1}{m_{\cal S}^2}\R\Delta^p \approx  \f{\sqrt{ \ld_{sx}}}{2\sqrt{2}}\f{m_p}{v_s}\sin2\theta \L\f{1}{m_{h_{\rm SM}}^2}-\f{1}{m_{\cal S}^2}\R\Delta^p, 
\end{align}
where, to get the final expression for $f_p$, we have used $m^2_{X}\approx \ld_{sx}v^2_s/2$ as a result of the DM mass gensis from CSISB.

There are several limits of special attention in studying the DM direct detection bound. First is the degenerate limit $m_{h_{\rm SM}}\approx m_{\cal S}$, which leads to a cancelation in the scalon portal contribution to $a_q$, manifest in Eq.~(\ref{scalon})~\footnote{This cancelation is not accident and not novel~\cite{Guo:2015lxa}. However, in the usual cases where all Higgs bosons gain masses from a tree-level potential, the more degenerate means the system is more mixed. In the SI setup, the scalon gets mass just at quantum level, allowing degeneracy without a large mixing angle.}; hence a relatively larger mixing angle is tolerated given a substantial degeneracy. But note that the Higgs portal contribution does not show  cancelation. Second is the heavy scalon limit $m^2_{\cal S}\gg m_{h_{\rm SM}}^2$, and then its contribution is suppressed compared to the SM Higgs boson. In other words, the scalon portal effectively becomes the SM Higgs-portal and one can not rely on a heavy scalon to suppress $\sigma_{\rm SI}$. It holds, of course, assuming that $m_{\cal S}^2$ and $\theta$ are totally independent. The third is the opposite limit with $m^2_{\cal S}\ll m_{h_{\rm SM}}^2$, which requires a small mixing angle to avoid the direct detection bound. To be more specific, considering a TeV scale DM, then we have the upper bound on the mixing angle,
\begin{align}
\sin\theta\lesssim 0.07\times \L\f{m_{\cal S}}{100\rm GeV}\R \L\f{m_X}{1{\rm TeV}}\R
 \L\f{\sigma^{\rm upper}_{\rm SI}}{10^{-9}{\rm pb}} \R^{1/2} \L\f{0.35}{\Delta^p}\R\rm GeV^{-1}. 
\end{align}
where we have used the estimation $m_{\cal S}^2\sim 10^{-2} \ld_{sx}^2 v_s^2$, but it may shows a sizable deviation by virtue of the mixing effect. In any case, heaviness helps to alleviate the stringent direct detection bound on the scalar DM~\cite{Gao:2011ka}. 


%

%
%


\subsection{Interplay between DM and radiative CSISB: Numerical results }

Now we numerically demonstrate the implications of the DM trigger to radiative CSISB, to figure out the viable parameter space giving both successful CSISB and DM. As argued, the Higgs portal coupling $\lambda_{hx}$ plays no dynamics roles, and moreover should be irrelevantly small, thus simply set to $10^{-3}$. Then there are four parameters $\lambda_h$, $\lambda_s$, $\lambda_{hs}$ and $\lambda_{sx}$ relevant to DM or/and CSISB. However, only one is free because of the three additional requirements for the correct weak scale, Higgs boson mass and DM relic density. The single free parameter makes the parameter exploration become easy. It is further restricted by two constraints: One is from the current LHC Higgs data, which sets the upper bound on the SM Higgs and scalon mixing angle $|\sin\theta|<0.44$~\cite{Loebbert:2018xsd}; the other one is from DM direct detection experiments such as XENON1T and PandaXII~\cite{Aprile:2017iyp}, which sets an upper bound on $\sigma_{\rm SI}$ for a given DM mass.

\subsubsection{Higgs-portal scenario}

We first investigate the allowed parameter space in the Higgs-portal scenario, where $\lambda_{sx}$ is chosen as the free parameter. We find that, even taking into account the large quantum effect, the heavier eigenstate $\phi_{+}$ in Eq.~(\ref{eq:aperp2.10}) cannot be identified with the SM-like Higgs boson. Let us briefly explain the reasons. To satisfy the conditions $m_{\phi_+}=125$ GeV and the correct DM relic density via Eq.~(\ref{relic}), the viable region is $0<\lambda_{sx}<1.04$. However, one meets $m_{\phi_-}^2<0$ for $\lambda_{sx}<0.33$ while $|\sin\theta|>0.44$ for $0.33<\lambda_{sx}<1.04$. So we have to identify the SM-like Higgs boson with the lighter one $\phi_{-}$. Then, for a given $\lambda_{sx}$, the solution of the equation $m_{\phi_-}=125$ GeV as a function of $\lambda_{hs}$ has two branches. One branch requires a large Higgs-scalon coupling $\lambda_{hs}>\lambda_{sx}$. But recall that the Higgs-portal limit requires a small $\lambda_{hs}$, so this branch is unacceptable. Whereas the other branch $\lambda_{hs}<\lambda_{sx}$ gives successful phenomenology if $\ld_{sx}$ is in the region $1.09<\lambda_{sx}<\pi$, with the upper bound for the sake of perturbativity. The condition $|\sin\theta|<0.44$ further shrinks the feasible parameter region to $1.16<\ld_{sx}<\pi$.

\begin{figure}[htbp] 
\centering 
\includegraphics[width=0.45\textwidth]{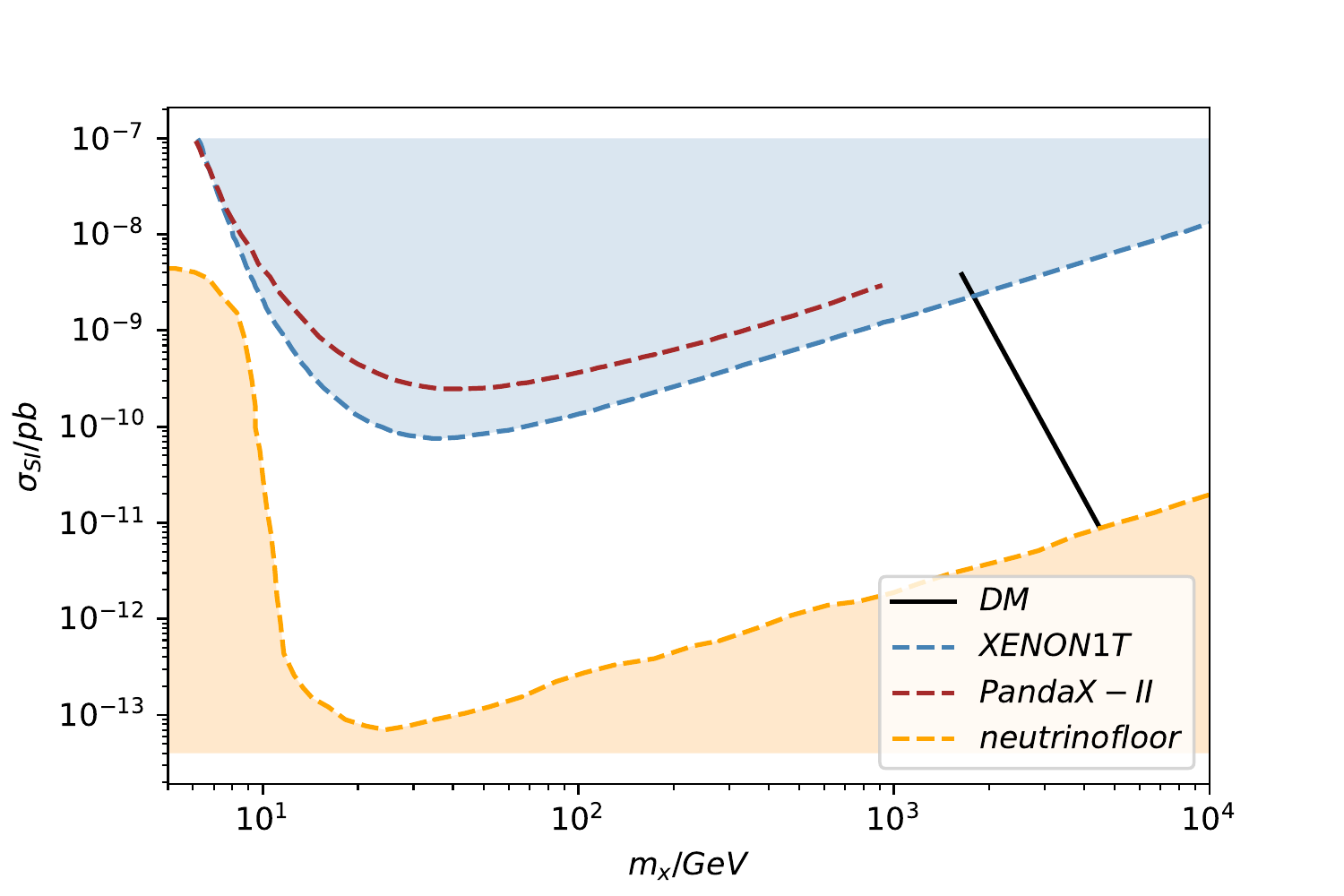} 
\includegraphics[width=0.45\textwidth]{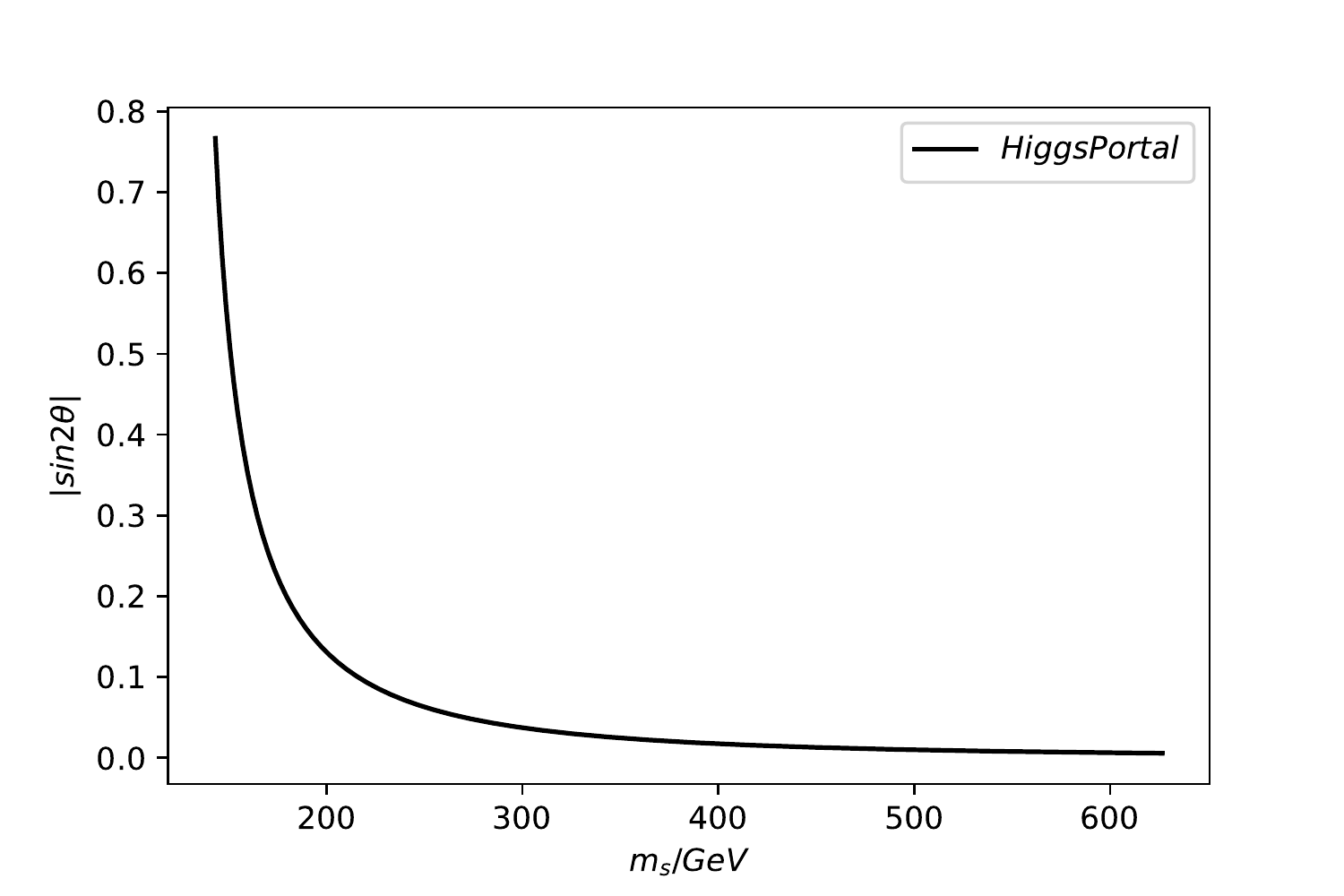} 
\caption{Left: The surviving parameter space under the DM direct detection bound (black line) in the Higgs-portal limit. Right: Profile of scalon in the  $m_{\cal S}-|\sin2\theta|$ plane for $0.46<\lambda_{sx}<\pi$.} \label{DD:Higgsportal}
\end{figure}
Now we add the stringent bound from DM direct detection. This scenario, by definition, should give a fairly small mixing angle. In the right panel of Fig.~\ref{DD:Higgsportal} we plot the profile of the scalon, on the $m_{\cal S}-\sin2\theta$ plane. One can see that $|\sin\theta|\lesssim {\cal O}(10^{-1})$ as long as the scalon mass is not very close to $m_{h_{\rm SM}}$, which is good for evading the DM direct detection bound for a relatively heavy scalon hence DM. This is just what the left panel of Fig.~\ref{DD:Higgsportal} shows:  DM with mass $m_X\gtrsim 1.78$ TeV is still allowed, corresponding to $\lambda_{sx}\gtrsim 1.26$ and a scalon with mass $m_{\cal S}\gtrsim 160$ GeV. In summary, in the Higgs-portal scenario the allowed parameter space lies in the interval:
\begin{align}
1.26<\lambda_{sx}<\pi.
\end{align}
Moreover, the resulting scalon, the clear prediction of the CSI models, tends to be heavy and slightly mixed with the SM Higgs boson thus hard to be probed at LHC.




\subsubsection{The GW scenario}\label{GWsc}

\begin{figure}[htbp] 
\centering 
\includegraphics[width=0.9\textwidth]{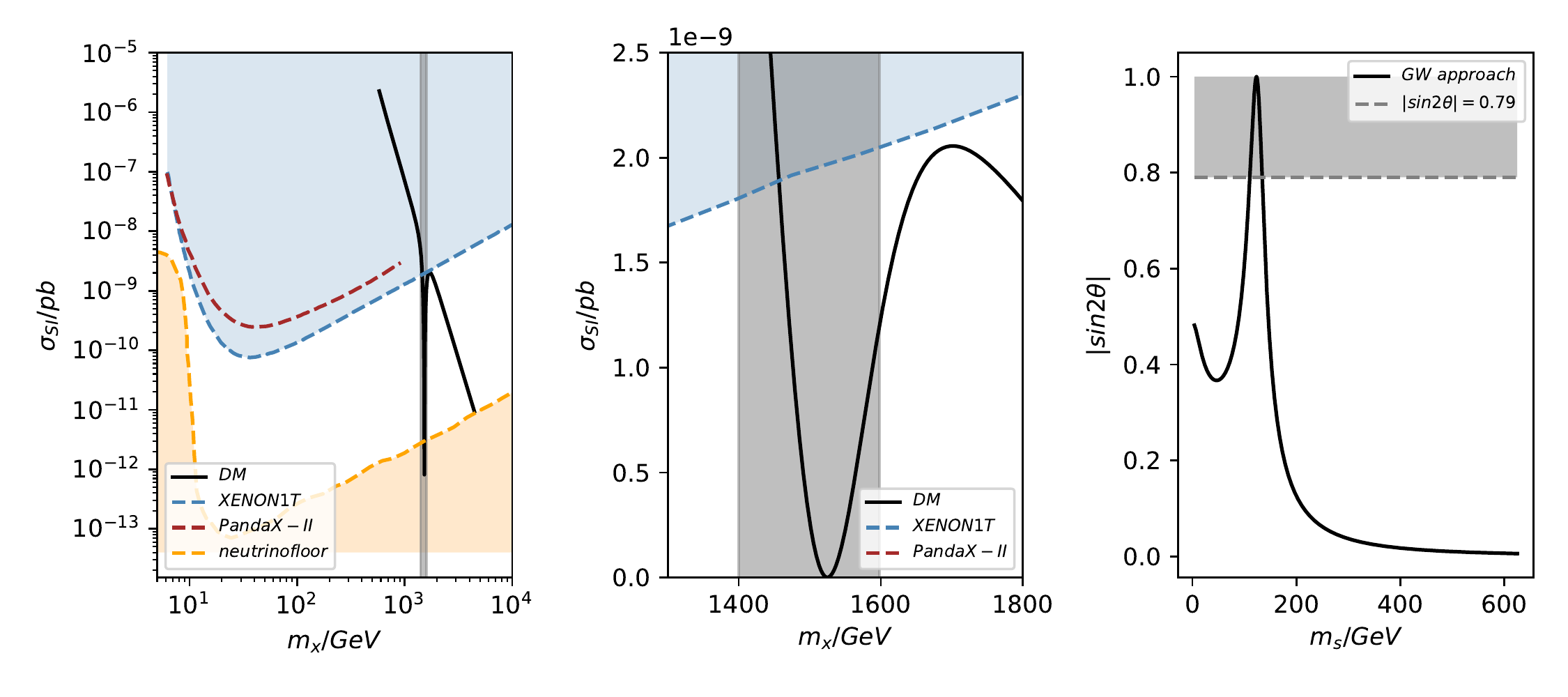}
\caption{Left: The surviving parameter space under the DM direct detection bound in the GW scenario; Middle: The zoom region showing cancelation; Right: Profile of the scalon in the $m_{\cal S}-|\sin2\theta|$ plane. The gray shaded region is excluded by Higgs data.} \label{Fig_CW}
\end{figure}

Next we study the GW scenario, where again $\lambda_{sx}$ is chosen as the free parameter. Then we add the constraints step by step. First of all, the condition for a stable vacuum namely $B>0$ sets the lower bound $\lambda_{sx}>0.23$. Next, to identify the 125 GeV SM-like Higgs boson with $\phi_{-}$ or $\phi_+$, it is found $\lambda_{sx}>1.08$ or $\lambda_{sx}<1.08$. Finally, the upper bound on the mixing angle $|\sin 2\theta|<0.79$ selects two regions $0.07<\lambda_{sx}<0.99$ or $1.13<\lambda_{sx}<\pi$. These constraints still allow a wide parameter space: 
\begin{itemize} 
\item  $0.23<\lambda_{sx}<0.99$ where $h_{\rm SM}$ is identified with the heavier one $\phi_{+}$ and DM mass is in the region $325 {\rm GeV}<m_{X}<1400 {\rm GeV}$.
\item $1.13<\lambda_{sx}<\pi$ where $h_{\rm SM}$ is the lighter one, $\phi_{-}$ and DM mass is in the region $1598 {\rm GeV}<m_{X}<4442 {\rm GeV}$ .
 \end{itemize} 
But $\sigma_{\rm SI}$ may be not sufficiently suppressed and consequently the DM direct detection significantly shrinks the viable regions; see the Fig.~\ref{Fig_CW}. The relatively light DM region has been excluded except for the narrow trough which shows subtle cancelation as discussed below Eq.~(\ref{sigmaSI});

 Only the relatively heavy DM mass region $m_X >1.56$ TeV survives, which indicates $\lambda_{sx} >1.03$ and $m_{\cal S} >117$ GeV. The scalon in the GW scenario is relatively light and moreover has a larger mixing angle, so it has better prospect at the LHC. 

\section{Supercooling CSI phase transition (CSIPT)}

Having studied CSI radiatively breaking at zero temperature, in this section we go back to the early universe with high temperature, where CSI is recovered. Of interest, it is found that the transition from the CSI phase to its broken phase is first order, usually characterized by a large supercooling~\cite{Witten:1980ez,Chiang:2017zbz,Marzo:2018nov,Prokopec:2018tnq,Ellis:2019oqb,Aoki:2019mlt,Mohamadnejad:2019vzg,Brdar:2019qut,Kubo:2016kpb}.

\subsection{Effective potential from finite temperature correction}

Cosmic PT is based on the finite temperature effect. When the background fields couple to a bath of plasma, its potential receives temperature dependent corrections from the thermal fluctuations of the plasma. The leading order finite temperature correction takes the following form~\cite{Quiros:1999jp},
\begin{equation} \label{finite} 
\begin{split}
V^{(1)}_T(\phi,T)=\frac{T^4}{2\pi^2}\left(\sum_{a\in boson}n_a J_B(x_a)+\sum_{a\in fermion}n_a J_F(x_a)\right),
\end{split}
\end{equation}
with $x_a={m_a(\phi)}/{T}$. The formalism applies to $\phi$ with multi-component. Like Eq.~(\ref{CWpot}), the index $a$ should run all heavy particles that couple to  the backgrounds, e.g., top quark and DM whose masses are given in Eq~(\ref{eq:aperp2.12},\ref{eq:aperp2.10}). When working in the Higgs portal scenario, we only need to include DM and scalon because other particles are massless at the stage of CSIPT. In particular, the absence of Higgs VEV simplifies the masses of scalon and DM to be
\begin{equation} \label{HiggsPortalMass} 
\begin{split}
m_s^2=b_X s^2,~~\  \  \  m_X^2=\f{\lambda_{sx}}{2}s^2.
\end{split}
\end{equation}
We will use those mass and potentials in following discussion.

In Eq.~(\ref{finite}), $J_B$ and $J_F$ are the thermal functions for bosons and fermions, and they respectively are given by
\begin{equation} \label{eq:aperp2.20} 
\begin{split}
J_{B/F}(y)= \int_0^{\infty}dx x^2 \log\left(1\mp e^{\sqrt{x^2+y^2}}\right).
\end{split}
\end{equation}
In the $y^2\ll 1$ limit, the above integrals admit the high temperature expansion, up to the quartic terms, taking the forms~\cite{Quiros:1999jp}
\begin{equation} \label{HT} 
\begin{split}
&J_B(y)\simeq -\frac{\pi^4}{24}+\frac{\pi^2}{12}y^2+\frac{\pi}{6}y^3-\frac{1}{32}y^4 \log\frac{y^2}{a_b}+{\cal O}(y^4)
\\
&J_F(y)\simeq \frac{7\pi^4}{360}-\frac{\pi^2}{24}y^2-\frac{1}{32}y^4 \log{\frac{y^2}{a_f}}+{\cal O}(y^4),
\end{split}
\end{equation}
with $\log a_b\approx 5.4$ and $\log a_f\approx 2.6$. One should be cautious about high temperature expansion in the PT with a large supercooling, where PT is completed at a very low $T$ and thus $y^2\ll 1$ does not hold. But this approximation is still adopted in some literatures to analyze such kind of PT,  just retaining the quadratic terms. We will come back to this point in the Section~\ref{results}, where we  argue how the expansion may still work.

The one-loop effective potential may be insufficient to describe PT. According to the principle that symmetry should restore at high temperature, the ordinary perturbative expansion in terms of coupling must break down at high temperature~\cite{symmetry:T,Dolan:1973qd}, e.g,, around or above the critical temperature $T_c$. To improve the expansion so as to make the analysis valid at high $T$, one should sum the high order diagrams which consist of the quadratically divergent loops on the top of the 1-loop self-energy diagram for the spin-0 particles. This procedure yields a thermal correction to their masses
\begin{equation} \label{} 
\begin{split}
M_a^2(\phi,T)=m_a^2(\phi)+\Pi_a(T),
\end{split}
\end{equation}
with $\Pi_a(T)$ specific to our model given by
\begin{equation} \label{self0} 
\begin{split}
&\Pi_h(T)=\Pi_{GSB}=\frac{\lambda}{4} T^2+\frac{\lambda_{hs}+\lambda_{hx}}{24} T^2+\frac{3g^2+g'^2}{16} T^2+\frac{y_t^2}{4} T^2,
\\
&\Pi_s(T)=\frac{\lambda_s}{4} T^2+\frac{\lambda_{hs}+\lambda_{sx}}{24} T^2, \ \ \ \ \ \ \ \ \Pi_X(T)=\frac{\lambda_x}{4} T^2+\frac{\lambda_{hx}+\lambda_{sx}}{24} T^2,
\end{split}
\end{equation}
and as well the longitudinal components of the gauge bosons
\begin{equation} \label{self1} 
\begin{split}
\Pi_{W_L}(T)=\frac{11}{6} g^2 T^2,\ \ \ \ \ \ \ \ \Pi_{Z_L}(T)\approx \frac{11}{6} (g^2+g'^2) T^2.
\end{split}
\end{equation}
We have neglected $\gamma_L$ which is not important numerically. Daisy resummation generates the daisy term in the effective potential~\cite{daisy,Curtin:2016urg}, 
\begin{equation} \label{eq:aperp2.23} 
\begin{split}
V_D(\phi,T)=-\frac{T}{12 \pi}\sum_{a\in boson} n_a\left([m_a(\phi)^2+\Pi_a(T)]^{\frac{3}{2}}-m_a(\phi)^3\right)
\end{split}
\end{equation}
where $a$ runs over the spin-0 fields and the longitudinal components of the gauge bosons that appear in Eq.~(\ref{self0}) and Eq.~(\ref{self1}) with $n_{W_L}=2, g_{Z_L}=1$.

\subsection{Bubble nucleation rate}

According to the work of Coleman and Callan~\cite{Vdecay},  first-order PT proceeds via the bubble nucleation of the true vacuum. The bubble nucleation rate per volume and per time $\Gamma(T)$, due to thermal fluctuations, is given by~\cite{Nuclrate}
\begin{equation} \label{} 
\begin{split}
\Gamma\approx A T^{4} e^{-\frac{S_3(T)}{T}},
\end{split}
\end{equation}
where $A$ is supposed to be at order 1. $S_3$ is the $O(3)$ symmetric three-dimensional Euclidean action 
\begin{equation} \label{S3} 
\begin{split}
S_3(T)=4\pi \int_0^{\infty} R^2dR \left[\frac{1}{2}\L\frac{d\phi}{dR}\R^2+V_{eff}(\phi,T)\right],
\end{split}
\end{equation}
with $R=\vec x^2$ and $V_{eff}(\phi,T)=V^{(0)}(\phi)+V^{(1)}(\phi)+V_T^{(1)}(\phi,T)+V_D(\phi,T)$ the total effective potential. $\phi(R)$ is the bounce solution satisfying the Euclidean equation of motion
\begin{equation} \label{} 
\begin{split}
\f{d^2\phi}{dR^2}+\f{2}{R}\f{d\phi}{dR}=V_{eff}',
\end{split}
\end{equation}
with the boundary conditions $\underset{R\ra \infty}{\lim}{ \phi(R)}=0$ (the false vacuum position) and $\f{d \phi(R)}{dR}|_{R\ra 0}=0$. The bounce solution connects the true vacuum and the false vacuum, with phase interface namely the bubble wall localized at $R=0$ and $R$ denotes the distance to the wall. The region $R>0 (<0)$ is in the symmetric (broken) phase.


Denote $S_3(T)/T$ as $S(T)$ hereafter. Finding $S(T)$ or essentially the bounce solution is the basis to discuss PT and as well the gravitational wave, however, it heavily relies on the numerical codes, e.g., the python program CosmoTransition~\cite{cosm}. We will assess the GW approach at $T\neq 0$ and the Witten's analytical approximation specific to CSIPT.

\subsubsection{Multi-field: Tunneling along the flat direction versus full tunneling}\label{GW:full}

\begin{figure}[htbp] 
\centering 
\includegraphics[width=0.9\textwidth]{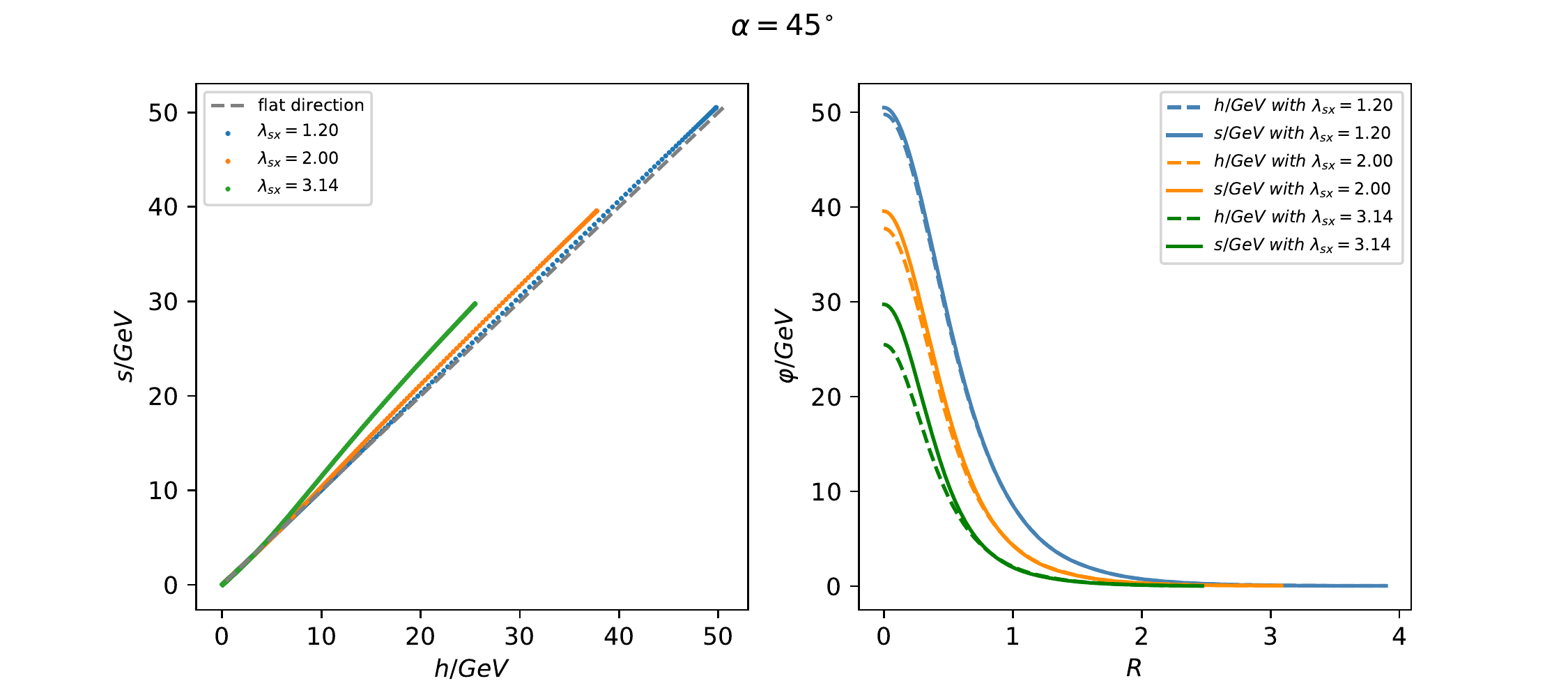} 
\includegraphics[width=0.9\textwidth]{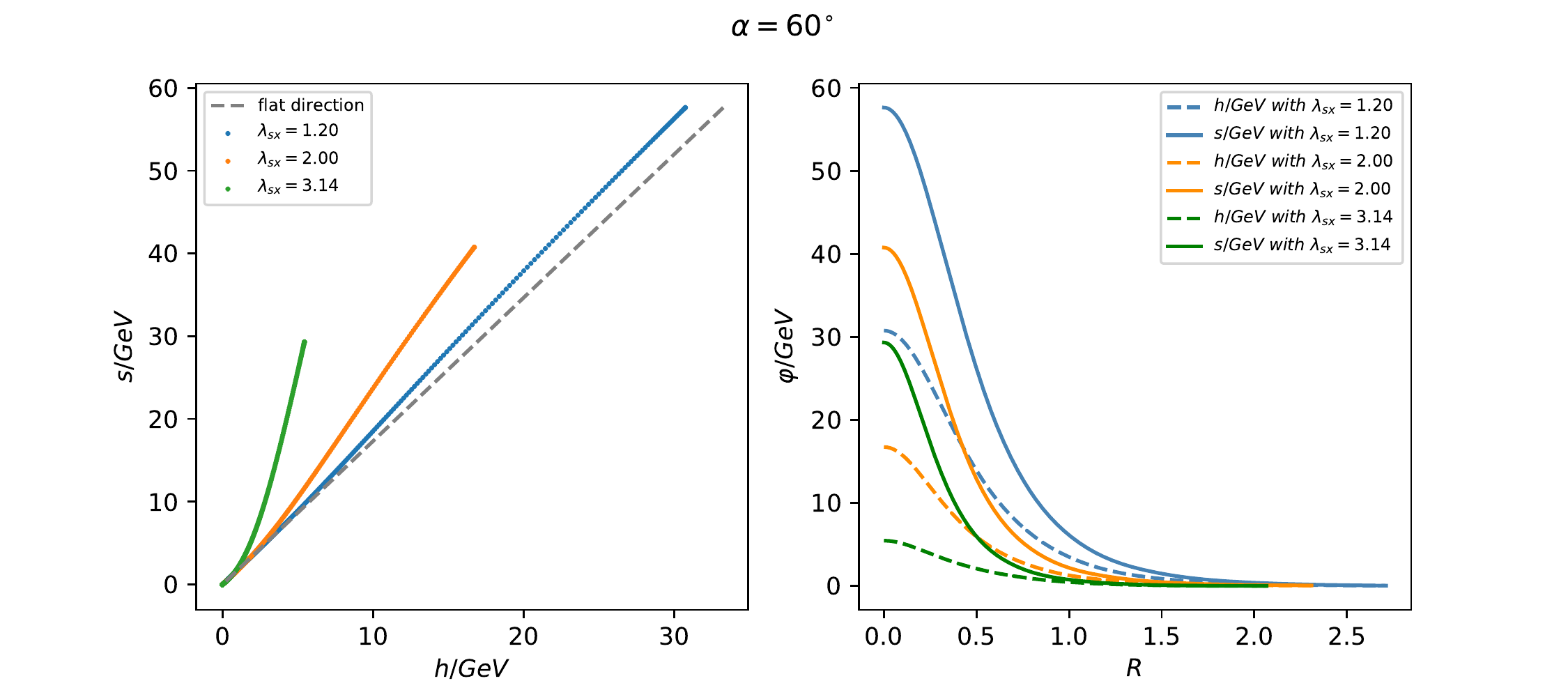} 
\caption{In this diagram $\alpha$ is the flat direction; all fields values are terminated around the escaping points, after which the particle follow the classical path and thus does not contribute to tunneling. Then, setting T=10GeV, we find the solution of tunneling problem with parameter $\lambda_{h}=0.13$, $\lambda_{hx}=0.001$ $\lambda_{x}=0.2$ and $v_h=246$ GeV. Other parameter is shown in this diagram or calculated by GW method. } \label{GWvality}
\end{figure}
In the GW approach dealing with radiative symmetry breaking in the multi-field space at $T=0$, analysis is done around the valley of the potential. While the calculation of finite temperature correction is also implemented along the flat direction (see Eq.~(\ref{FT:GW})), which means that we are assuming that thermal tunneling between the vacua  is along this direction. However, a strong quantum correction, present in the case of a larger coupling, may strongly distort the shape of the valley, and therefore we may wonder if the actual tunneling still follows the  flat direction. To that end, we study a few example points without considering any phenomenological constraints and the results are shown in Fig.~\ref{GWvality}. Two cases of flat direction are presented: One is along $h=s$ while the other one is along $h=s/\sqrt{3}$. From the left panels one can see that, as expected, the tunneling path begins to deviate away from the flat direction significantly as $\ld_{sx}$ thus quantum correction increases. Note that for a given $\ld_{sx}$, quantum correction leads to a larger deviation for the case with a larger $s$, because the correction mainly comes from the trigger-$s$ coupling. In fact, given that $s\gg h$ holds during tunneling, the multi-field problem effectively is reduced to the single-field problem in the sense of calculating $S(T)$.

\begin{figure}[htbp] 
\centering 
\includegraphics[width=0.9\textwidth]{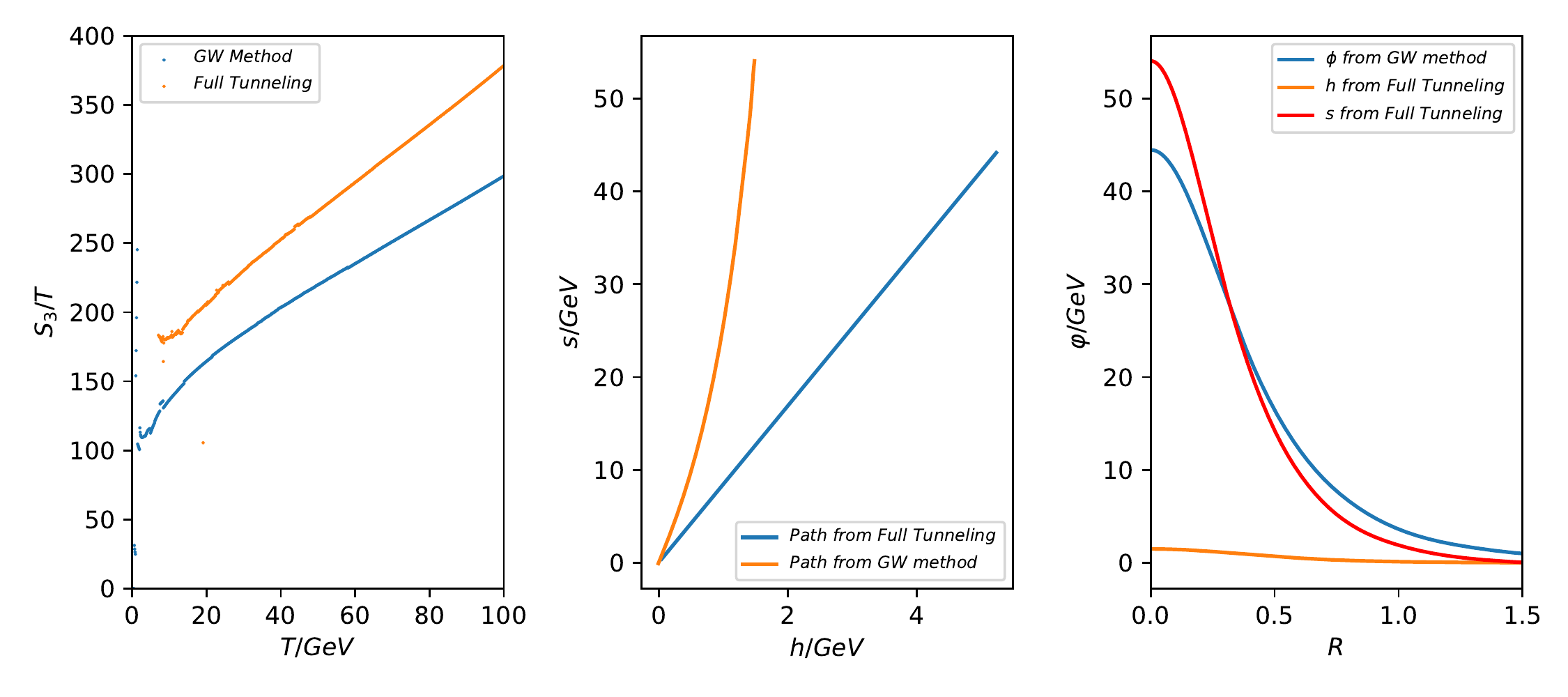} 
\caption{Left panel: Comparison of $S(T)$ between the calculation along the flat direction (top) and along the actual path (bottom).Middle panel:  The tunneling path in the $s-h$ space.  Right panel: The tunneling path in the $R-\phi$ space. The parameter set is $\lambda_{sx}=1.2$, $\lambda_{h}=0.1273$, $\lambda_{s}=0.000025$ and $v_s=2076\rm GeV$.} \label{S3:GW}
\end{figure}

However, $S(T)$ is not  very sensitive to the tunneling path but sensitive to the position of the escaping point. To show this we calculate  $S(T)$ for  $\lambda_{sx}=1.2, 2.0$ and $3.14$,  using both the GW approach, where tunneling faithfully tracks the flat direction, and the multi-field full tunneling, where tunneling is along the actual trajectory, to get
\begin{equation} \label{S3:com} 
S(T={\rm 10 GeV})= 387 (335), 139 (138)~{\rm and}~~ 50 (62), 
\end{equation}
respectively; values in the brackets are for the full tunneling. Analysis on the variation of $S(T)$ with $\ld_{sx}$ will be given in Section.~\ref{results}. The above examples indicate that the difference between the two ways is mild, typically below 20 percent. Moreover, it seems that the degree of difference has no simple correlation with the degree of path deviation. In summary, tracking the tree-level flat direction still provides an acceptable approximation to study PT, even facing a relatively strong  quantum correction. This conclusion is further supported by a realistic example in Fig.~\ref{S3:GW}, and its first diagram is a  comparison between $S(T)$ from two approaches in a wide region of temperature.

\subsubsection{Single field: Witten's approximation?}\label{WA}

If the CSIPT involves only one scalar field, the Witten's approximation is usually adopted to estimate $S(T)$ at the very low $T$ region~\cite{Witten:1980ez}, for instance in the conformal local $B-L$ model~\cite{Jinno:2016knw}. Witten observed that for very low $T$ the field contributing to tunneling extends to $\phi\sim T/\ld$ with $\ld$ denoting the coupling between $\phi$ and trigger~\footnote{One may proves it directly from the 1-loop effective potential, by finding its zero points, with one at the origin the metalstable vacuum while the other one the escaping point.}. This fact allows one to take high temperature expansion to derive the tunneling potential merely describing the tunneling process; it is approximated to be 
\begin{equation} \label{witten} 
V_{tun}(\phi,T)=\frac{m_{eff}^2(T)}{2}\phi^2+\frac{\lambda_{eff}(T)}{4}\phi^4,
\end{equation}
where $m_{eff}^2(T)$ is the effective mass in high temperature expansion and $\lambda_{eff}(T)$ is negative at low $T$. The vacuum decay of such a potential has been studied in Ref.~\cite{negative}, giving an analytical expression of 
\begin{equation} \label{formula} 
S(T)\approx -18.897\frac{m_{eff}(T)}{T \lambda_{eff}(T)}.
\end{equation}

Nevertheless, the Witten's approximation scheme just gives an estimation on $S(T)$ at very low $T$, and 
it is not good in the sense of precision. Here are two reasons:
\begin{itemize} 
\item First, in the original treatment, only the quadratic term is kept in the high tempera-
ture expansion. In particular, the cubic term $\phi^3$, which plays an important role in the
shape of the barrier, is simply dropped; on the other hand, keeping this term one can
not write the tunneling potential in the form of Eq.(\ref{witten}). Such over simplification
gives rise to a significant deviation to the complete result. If one includes the daisy
term (it is not also included in the original paper), it will exactly
cancel that $\phi^3$ term, but leaving the cubic term of the thermally corrected
trigger mass.
\item Second,  the crucial negative quartic coupling is not unique because it is derived by a rough argument rather from first principle:  Around the escaping point  $\phi\sim T/\ld$, the logarithmic term in the  CW potential $ \log \f{\phi}{\mu}= \ln \f{ T}{m}+ \ln \f{\ld\phi}{ T}\sim  \ln \f{ T}{m}<0$, where $m\simeq \lambda \mu$ is the physical mass of the trigger. The drop of the $\ln \f{\ld\phi}{ T}$ term is  justified in the very small $\phi\ll T$ region, because the quartic term is irrelevant. But obviously one has some degree of arbitrariness to split $\log\f{\phi}{\mu}$. Actually, a similar expression can be derived if we keep terms up to the quartic term in the high temperature expansion. This quartic term and the quartic term in the CW potential have similar coefficients, and they combine to form 
\begin{equation} \label{} 
\log\f{ a_b T^2}{\phi^2}+\log\f{\phi^2}{\mu^2}=\log \f{a_b T^2}{\mu^2}.
\end{equation}
So, the negative quartic coupling is derived without turning to $\phi\sim T/\ld$.
 \end{itemize} 
In the above discussions we actually modify the Witten's approximation, maintaining the high temperature expansion (to quartic terms) but giving up the formula Eq.~(\ref{formula}).

\begin{figure}[htbp] 
\centering 
\includegraphics[width=0.75\textwidth]{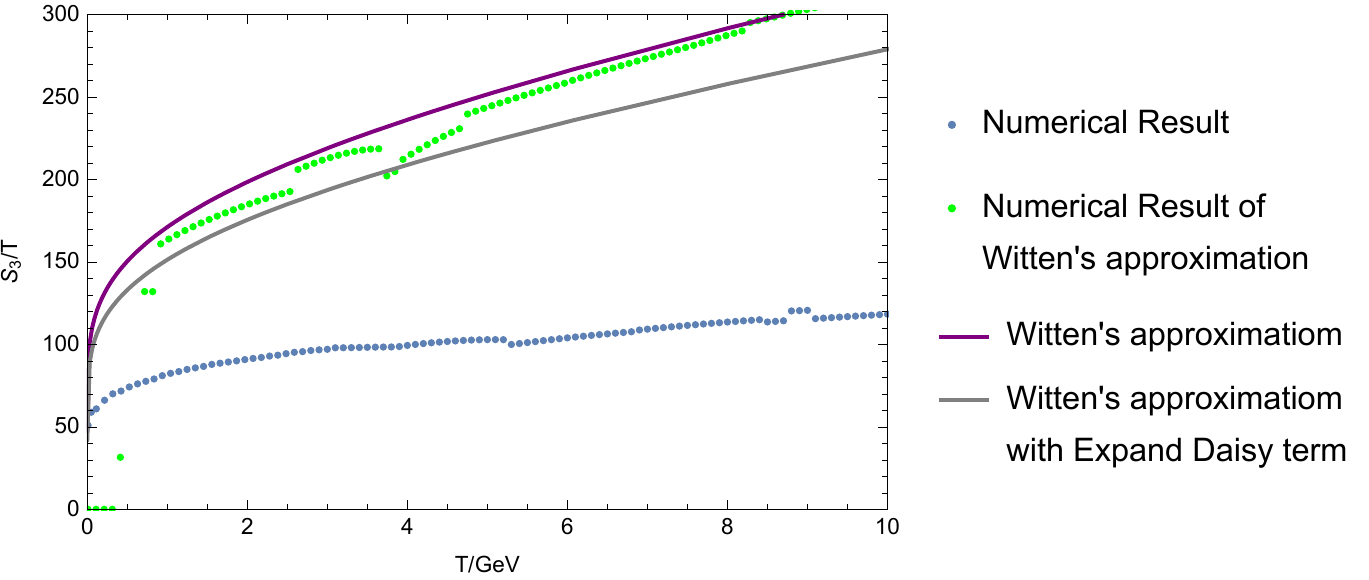} 
\caption{Various Witten's approximation versus the numerical results: The numerical results for the full potential (bottom dotted line) and for the tunneling potential from the original Witten's approximation  (dotted green line), which indeed is almost the same  with its analytical result  (the top line); Witten's approximation for the potential after expanding the daisy term as in Eq.~(\ref{daisy:exp}) (gray solid line).} \label{witten:app}
\end{figure}

To be more specific, we apply the modified  Witten's approximation to our model in the hidden CW scenario, only  taking into account the DM field $X$. First of all, high temperature expansion indeed works well: In  Fig.~\ref{witten:app}  the blue dotted line denotes the numerical result of the complete potential, and it  well coincides with the line (not plotted explicitly) for the potential in high temperature expansion. Then, we  derive the tunneling potential from the high temperature expansion,
\begin{equation} \label{fulldaisy} 
\begin{split}
V_{tun}(\phi,T)
&=\f{1}{4}\lambda_{s}\phi^4+\frac{1}{64\pi^2}m_{X}^4(\phi)\left(\log\frac{a_b T^2}{\mu^2}-\frac{3}{2}\right)+\frac{m_{X}^2(\phi) T^2}{24}-\frac{M_{X}^3(\phi,T) T}{12\pi}.
\end{split}
\end{equation}
As mentioned before, the term $M^3_X(\phi,T)\approx (\ld_{sx}/2)^{3/2}(\phi^2+T^2/12)^{3/2}$ hampers the direct using of Witten's formula. Hence we further expand it in terms of $\phi/T$, up to the quadratic term, and then the tunneling potential takes the form of Eq.~(\ref{witten}) with
\begin{equation} \label{daisy:exp} 
\begin{split}
&m_{eff}^2(T)=\f{m_X^2(T)}{12}-\f{m_X^2(T)}{4\pi}\sqrt{\f{6\lambda_s+\lambda_{sx}}{24}},
\\
&\lambda_{eff}(T)=\lambda_s+\f{\lambda_{sx}^2}{64\pi^2}\left(\log\frac{a_b T^2}{\mu^2}-\frac{3}{2}\right).
\end{split}
\end{equation}
To check if this approximation works well, we compare the resulting $S(T)$ by Witten's formula with the complete numerical results, to find that it is a poor approximation; see the Fig.~\ref{witten:app}. This inaccuracy is owing to the fact that the expansion $\phi/T\sim {\cal O}(1)$ is multiplied by a large factor $\simeq 12\ld_{sx}\gg 1$. Therefore, we draw the conclusion that Witten's formula does not give a precise estimation on $S(T)$.  

\subsubsection{Generalized Witten's argument?}

Maybe the essence of the Witten's argument is not the poor formula for estimating $S(T)$, but the observation that high temperature expansion is a good approximation to encode the tunneling dynamics far below the critical temperature. More concretely, the barrier, in particular the escaping point, just extends over the small field region and thus the quantum tunneling path merely tracks small fields. Whereas the large fields, where the ground state is located, are irrelevant. This fact, along with the CSI, confers the legitimacy of the high temperature expansion at very low temperature.

\begin{figure}[htbp] 
\centering 
\includegraphics[width=0.45\textwidth]{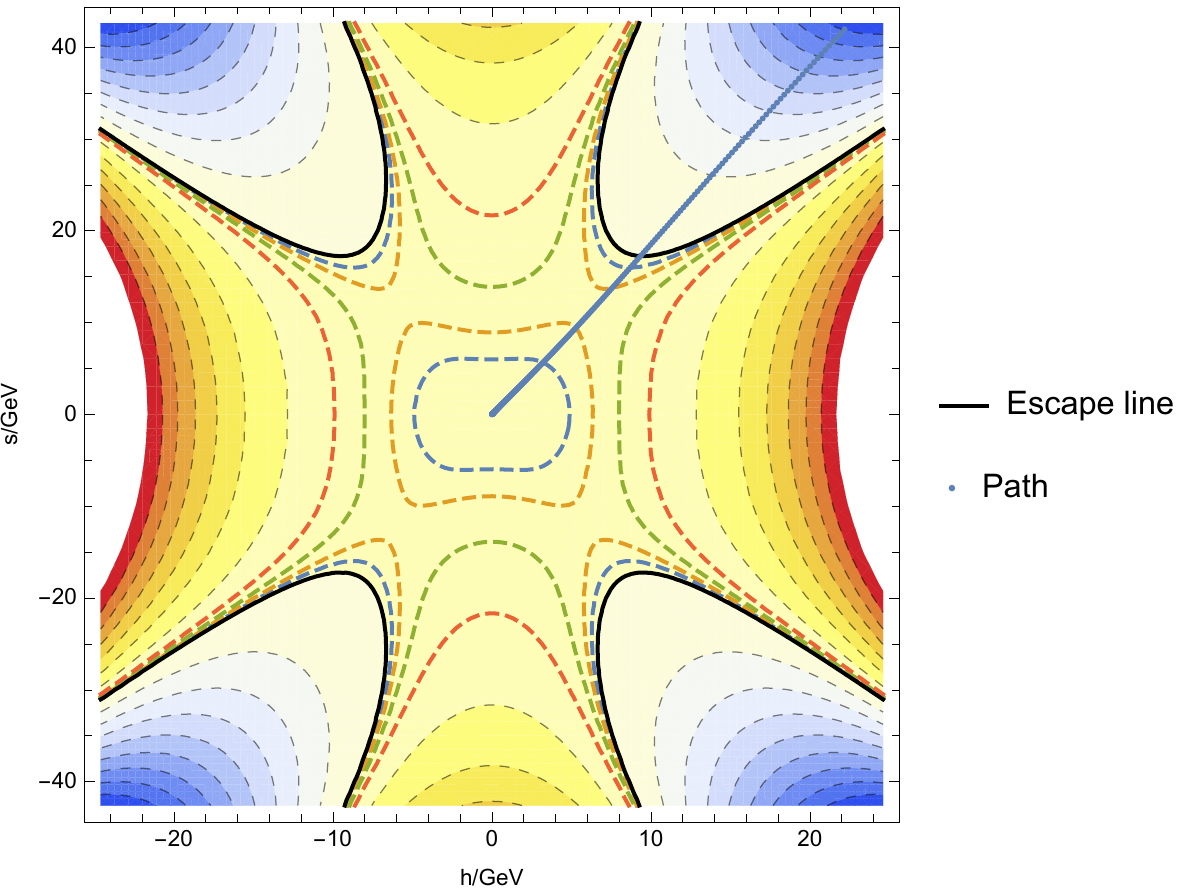} 
\includegraphics[width=0.45\textwidth]{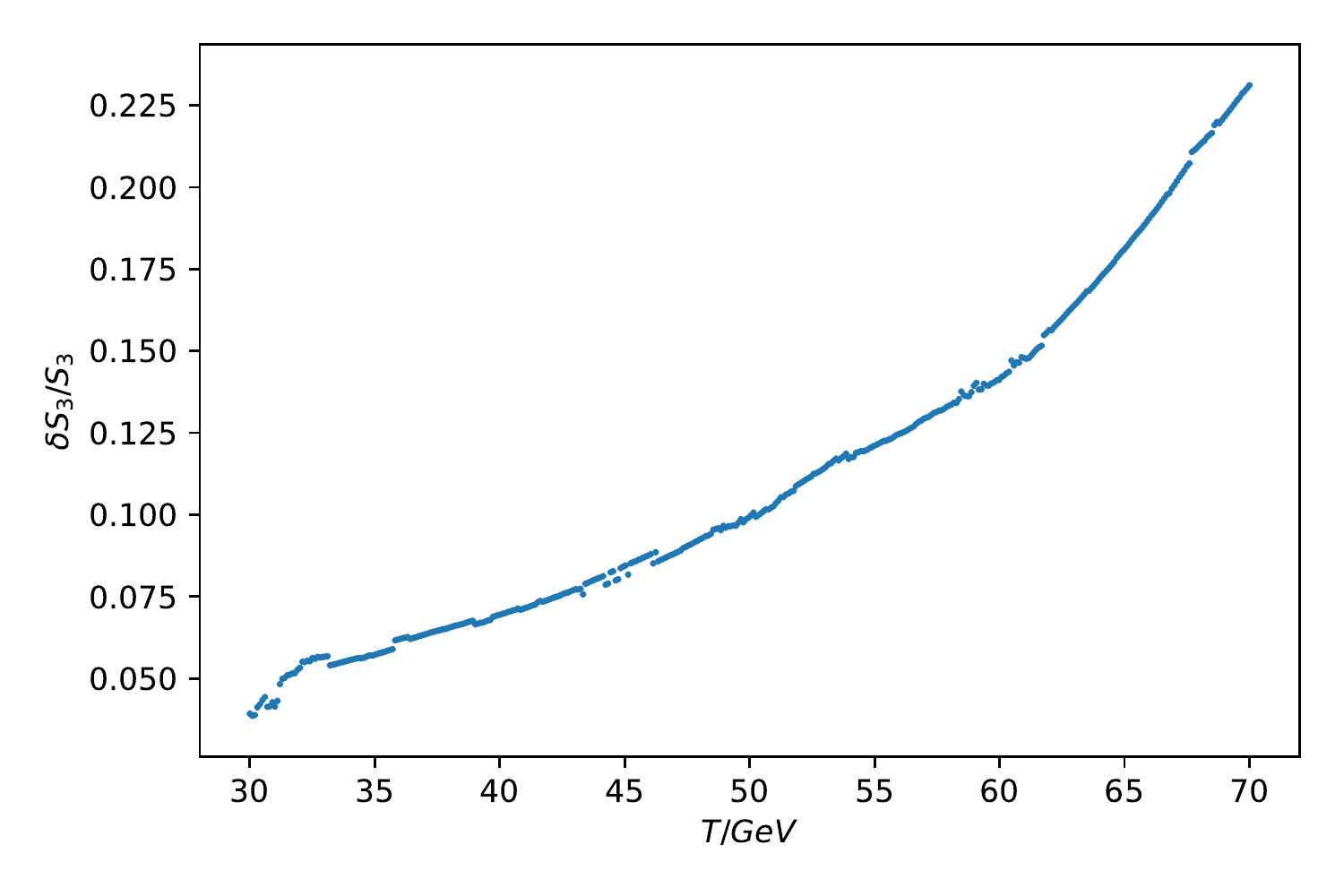} 
\caption{Left panel: Contours of equipotential lines of the model and the tunneling path (straight line). Right panel: The relative error of $S(T)$ for high temperature expansion.} \label{MF}
\end{figure}

The original argument is for the single field, and we conjecture that it may also applies to the multi-field case. But it is difficult to prove it explicitly since, unlike the 1-dimensional case, the escaping points now are located in a hypersurface in the $n$-dimensional field space and we are incapable of pinning down the exact point at which the tunneling ends. Moreover, the scale of fields varies widely on the escaping hypersurface, which renders the failure of the simple conclusion that the tunneling process just involves fields extending to $T/\ld$. But we conjecture it is true. A support is from the left panel of Fig.~\ref{MF}, where the contours are the equipotential lines of our model, and the thick black line with zero potential energy is the escaping line. Its interaction with the straight line, the tunneling path, is the actual escaping point. Thus as our conjecture the tunneling is through the small field region and ends at the small field, and then the Witten's argument is supposed to hold. 

Even if the above generalization is true, to compute $S(T)$ we still have to rely on the numerical codes. Our discussions help to clarify what is the correct way of using the high temperature expansion in CSIPT; some authors merely keep the quadratic terms of $T$, but it does not give good enough numerical results; more details can be found in Appendix.~\ref{AA}. In order to have a sufficiently good result, we need to expand the finite temperature potential to the quartic terms. In the right panel of Fig.~\ref{MF}, we show the quality of high temperature expansion in this scheme, measured by $\delta S(T)/S(T)$: the difference between the $S(T)$ calculated using the complete potential and the expanded one, normalized by the complete result. We can see that the quality is steadily improved as temperature decreases, contrast to the behavior of normal high temperature expansion.

\subsection{CSIPT in a hot bath or in the vacuum?}

As a consequence of very strong supercooling CSIPT, the early universe experienced a very short stage of vacuum dominated era, thus a short period of little inflation. CSIPT may be completed during this epoch rather than the usual radiation dominated (RD) era, and then we should reconsider the condition of CSIPT completion, which is recently stressed in Ref.~\cite{Ellis:2018mja,Ellis:2019oqb}. This is not very new, and the discussion is a reminiscence of the old inflation idea proposed by Guth~\cite{oldinflation}, but here the little inflation will be ended by thermal instead of quantum tunneling. 

%
%
%


\subsubsection{Little inflation}

In the scenario of supercooling PT, the universe was confined in the false vacuum until the PT completion temperature $T_*$, which lies much below the critical temperature $T_c$. Then, in this vacuum the nonvanishing vacuum energy density $\rho_0$, nearly a constant not diluting (we will come back to this point soon later), may began to exceed the radiation energy density $\rho_r(T)=\f{ \pi^2}{30}g_* T^4$ at some lower temperature $T_{V} \simeq \L 30 g_*\rho_0/\pi^2\R^{1/4}$, with $g_*\sim100$ the relativistic degrees of freedom in the false vacuum plasma. 

$\rho_0$ is determined by the potential energy of the false vacuum. The effective scalar potential has $T$-dependence, and therefore in principle $\rho_0$ also depends on $T$. However, since we are interested in the region near $T_*$, which is low due to supercooling, the finite temperature effect becomes fairly weak. Roughly speaking, this effect merely reshapes the potential near the origin (the small field region), maintaining the local minimum; it does not significantly change the ground state (the relatively large field region). So, it is a good approximation to calculate the vacuum energy from the effective energy at $T=0$~\footnote{The true vacuum energy should be fine-tuned to be zero by adding a constant to the potential, which is the usual cosmological constant problem. This constant is not dimensionless thus explicitly breaking CSI. We do not have an approach to reconcile CSI with it in this paper.}: 
\begin{equation} \label{VG} 
\begin{split}
 \rho_{0}= V_0^{(1)}(0,T_n)-V_0^{(1)}(\left\langle\phi\right\rangle,T_n)=\frac{1}{2}B \left\langle\phi\right\rangle^{4}.
\end{split}
\end{equation}
It is for the GW scenario, and a similar result can be derived in the Higgs portal scenario. As a result of CSI, its scale is mainly determined by the position of the ground state. Then, $T_{V}$ is estimated to be
\begin{equation} \label{} 
\begin{split}
T_{V} \simeq \L\f{15}{\pi^2}\f{B}{g_*}\R^{\frac{1}{4}} v_\phi. 
\end{split}
\end{equation}
In this paper we are considering $v_\phi$ at the TeV scale while the prefactor $\sim 0.1$ given a normal loop function $B\sim 10^{-2}$, thus typically $T_V\sim 100$ GeV.

After the universe energy density is dominated by vacuum energy, the size of the universe grows exponentially by means of inflation,
\begin{equation} 
\begin{split}
a(t)=a_V e^{H_V(t-t_V)}, 
\end{split}
\end{equation}
where $H_V\approx \rho_0^{1/2}/(\sqrt{3}M_{\rm Pl})$ is the Hubble parameter during the vacuum dominated era; $t_V$ and $a_V$ are the time and scale factor at $T_V$, respectively. As in the inflation, we denote the Hubble times of $1/H_V$ as the e-folding number $N$: $t_N-t_V=N/H_V$. Here we consider the inflation with a smaller $N$ of a few, thus the little inflation. The temperature of the radiation drops exponentially, $T(t)=T_V  e^{-H_V(t-t_V)}$, and for $T_V\sim 100$ GeV the universe cools down to the sub-GeV after about $N\sim6$ Hubble times. If the CSIPT fails to complete before it, the QCD chiral PT will terminate inflation around this temperature~\cite{QCDPT}. In this paper we focus on the case that CSIPT is capable of ending inflation.~\footnote{Baryon asymmetry may be an issue if $N$ is very large.} In the following we investigate the condition for a successful CSIPT.

\subsubsection{ Condition for CSIPT completion }\label{criteria}

Despite of the difficulty to get an analytical expression for $\Gamma(T)$, practically it is sufficient to be aware of such a fact: In general, $\Gamma(T)$ monotonically decreases with $T$ since $S(T)$ increases with $T$. Therefore, the integration involving $\Gamma(T)$ is supposed to be dominated by the lower bound (in some sense, insensitive to UV). Then one has the following useful approximation,
\begin{equation} \label{app}
\begin{split}
\int^{T_c}_T \Gamma(T')T'^{n-4}dT'\approx  \int^{T_c}_T A e^{-\beta_0 T'}T'^{n} dT'\approx A\beta_0^{-n-1}e^{-S(T_0)}\Gamma(n+1,\beta_0 T),
\end{split}
\end{equation}
where we have expanded $S(T)$ around some temperature $T_0$: $S(T)=S(T_0)+\beta_0(T-T_0)+...$, retaining only the linear term. Note that $\beta_0\equiv dS(T)/dT|_{T_0}>0$. This treatment works very well for $T$ sufficiently close to $T_0$ and as well $T_c\gg T\sim T_0$. As a matter of fact, we will study $S(T_0)\sim {\cal O}(10)$, so it always works.

When does the bubbles of true vacuum overwhelmingly occupy the space of false vacuum? We label this temperature (time) as $T_n (t_n)$, known as the bubble nucleation temperature (time). In the RD epoch, the criterion is that at $T_n(t_n)$ a single bubble is nucleated within one Hubble horizon volume,
\begin{equation} \label{nuc} 
\begin{split}
N_n=\int_{t_c}^{t_{n}} dt \f{\Gamma(t)}{H(t)^3}=A
\int_{T_n}^{T_{c}}\frac{dT}{T^5}\L3{M^2_{Pl}}\R^2\L\f{30}{\pi^2 g_*}\R^2 e^{-S(T)} \sim 1. 
\end{split}
\end{equation}
To perform the integration over temperature, which is more convenient in PT, we have utilized the expansion rate in the RD epoch, $H(T)^2=\f{\pi^2 g_*}{30}T^4/(3M_{\rm Pl}^2)$ with $M_{\rm Pl}=2.43\times 10^{18}$ GeV and as well as the time-temperature relation 
\begin{equation} \label{} 
\begin{split}
dt/dT=-1/(HT).
\end{split}
\end{equation}
It holds for the universe evolving adiabatically, true both in the radiation and vacuum dominated era considered in this paper. Then following Eq.~(\ref{app}), the condition Eq.~(\ref{nuc}) is translated to the well-known equation,
\begin{equation} \label{Tn} 
\begin{split}
S(T_n) \simeq 
2\log (3M_{\rm Pl}^2/T_n^2)+2\log \f{15}{\sqrt{6}\pi^2 g_*}+\log f_R(x)\sim 140,
\end{split}
\end{equation}
where $f_R(x)=-6 + 2 x - x^2 + x^3 + x^4 e^x{\rm Ei}(x)$ with $x\equiv \beta_n T_n\sim{\cal O}(1)$ in our samples of numerical calculations.

But $S(T_n)\sim 140$ significantly overestimates the required value of $S(T_n)$ in the vacuum dominated epoch. Estimates Eq.~(\ref{nuc}) in this epoch thus $H=H_V$, the condition Eq.~(\ref{Tn}) turns out to be
\begin{equation} \label{Tn:VD} 
\begin{split}
S(T_n) \simeq 
2\log (3M_{\rm Pl}^2/T_n^2)+2\log {T_n^4}/{\rho_0}+\log f_V(x), 
\end{split}
\end{equation}
with $f_V(x)=(6 + 6 x + 3 x^2 + x^3)/x^4$. It is similar to the usual nucleation condition Eq.~(\ref{Tn}), but the term $2\log \rho_0$, originating in vacuum dominance, brings a significant numerical difference; now typically $S(T_n)\sim 70$. The concrete value of $\beta_n$, found to be $\sim {\rm GeV}^{-1}$ for a wide region of temperature in our model (which indicates that $S(T)$ is almost linear in $T$), is almost irrelevant in calculating $S(T_n)$.

The above bubble nucleation condition does not reflect the progress of PT, so one may develop a more apparent criterion via $P(t)$, the probability of a space point staying in the false vacuum~\cite{PT2}. The  criterion $P(t)\lesssim 70\%$ is usually used to fulfill percolation in the three-dimensional Euclidean space~\footnote{Successful bubble percolation is required to make the space homogeneous, namely the bubble do not form finite clusters. It is a more strict condition for PT completion.}. $P(t)=e^{-I(t)}$ with $I(t)$ the expected volume of true-vacuum bubbles per unit volume of space at time $t$~\cite{I(t)}, explicitly
\begin{equation} \label{} 
\begin{split}
I(t)=\int_{t_c}^t dt' \Gamma(t') a(t')^3V(t,t'),\quad V(t,t')= \f{4\pi}{3}r(t,t')^3,
\end{split}
\end{equation}
with 
\begin{equation} \label{} 
\begin{split}
r(t,t')=\int_{t'}^t v_w(t'') \f{dt''}{a(t'')}
\end{split}
\end{equation}
the comoving radius of the bubble nucleated at $t'$ expanding with a velocity $v_w$ until $t$. But $P(T)$ alone may be insufficient to judge if the PT is completed in the vacuum dominated era, where the $P(T)$ can be arbitrarily small nevertheless PT is never completed because of the inflation of the false vacuum~\cite{PT1}.

In such a case, a better condition for successful PT completion is obtained by finding the time $T_e$ since which the physical volume of the false vacuum ${\cal V}_{f}(T)=a(T)^3 P(T)$ commences to shrink~\cite{PT2}. It leads to the following condition 
\begin{equation} \label{shrink} 
\begin{split}
\f{1}{{\cal V}_{f}(t)}\f{d{{\cal V}_{f}(t)}}{dt}=3H(t)-\f{dI(t)}{dt}=H(T)\L3+T\f{dI(T)}{dT}\R\leq 0.
\end{split}
\end{equation}
For further analyzation we should pursue an approximation to $I(t)$. Because the bubble is very energetic in the very strong PT, it is safe to take $v_w(t)\approx 1$. Then, utilizing Eq.~(\ref{app}) and working in the vacuum dominated era, one can get~\footnote{In the RD era $I_{RD}(T)\approx \f{\pi }{18 H_R(T)^4} \Gamma(T) F(\beta_0 T)$ with $F(x)= \left(x^3+12 x^2+36 x+24\right) x e^x \text{Ei}(-x)+x^3+11 x^2+26 x+6>0$.}
\begin{equation} \label{ITVD} 
\begin{split}
I(T)\approx A\f{8\pi }{\beta_0^4 H_V^4}\exp\left[-S(T_0)-\beta_0(T-T_0)\right].
\end{split}
\end{equation}
Now, saturating the equality Eq.~(\ref{shrink}) and taking advantage of Eq.~(\ref{ITVD}) yields the equation $3-T_e\beta_e I(T_e)=0$, more concretely the PT completion condition
\begin{equation} \label{} 
\begin{split}
3-\f{8\pi e^{-S(T_e)}T_e}{\beta_e^3 H_V^4}=0\Rightarrow S(T_e)=2\log \f{3M_{\rm Pl}^2}{T_e^2}+2\log\f{T_e^4}{\rho_0}+\log \f{8\pi }{3x^3}.
\end{split}
\end{equation}
It is almost identical to Eq.~(\ref{Tn:VD}) except for the last term that is subdominant; actually, the difference is just a few for a widely changing $x$. Therefore, the difference is not sizable no matter using which criteria to measure the completion of PT.

The real implication of the latter criterion is that it forces ${\cal V}_{f}(T_e)$ to reach a maximum at $T_e$, and thus the second term of its Taylor expansion
\begin{equation} \label{} 
\begin{split}
{\cal V}_{f}(T)={\cal V}_{f}(T_e)\L1+\f{(T-T_e)^2}{2}\L3/T^2-\f{d^2I}{dT^2}\R|_{T=T_e}+...\R
\end{split}
\end{equation}
should have a negative coefficient at $T_e$. It contains two competitive pieces. One is from the curvature of $P(T)$, generating the native piece $-{d^2I}/{dT^2}|_{T=T_e}=-\beta_e^2 I(T_e)=-3\beta_e/T_e<0$. The other one is from volume expansion $a^3$, generating the positive piece $3/T_e^2$. The two pieces add up to a negative coefficient imposing a lower bound on the PT completion temperature, $T_e>3/\beta_e$. It is more convenient to rewrite the condition in terms of $\wt \beta$ that will be defined in Eq.~(\ref{betat})
\begin{equation} \label{cri:3} 
\begin{split}
T_e \beta_e=\wt\beta>3.
\end{split}
\end{equation}
By contrast, if PT completes in the RD era, there is no such kind of bound because $(3/T^2-dI^2/dT^2)|_{T_e}=3/T_e^2 {\cal F}(x)$ with ${\cal F}(x)$ definitely negative.



\subsection{Numerical results}\label{results}

\begin{table}  
\caption{ Benchmark points in the GW scenario}  \label{Tn:GW}
\begin{tabular*}{14.85cm}{|p{0.5cm}|p{1.5cm}|p{1.1cm}|p{1.7cm}|p{1cm}|p{1cm}|p{1cm}|p{1.3cm}|p{1.4cm}|p{1.4cm}|p{1.4cm}|}  
\hline  
 & $v_s/GeV$&$\lambda$&$ \lambda_s$&$ \lambda_x$&$\lambda_{hx}$&$\lambda_{sx}$&$\alpha$&$\wt \beta$&$T_n$/GeV&$T_{*}$/GeV\\  
\hline  
A &2400&0.1277&0.000014&0.2&$10^{-3}$&1.44&$3.9*10^8$&11.3&1.01&616 \\ 
\hline  
B &2449&0.1278&0.000013 &0.2&$10^{-3}$&1.50&$6.6*10^7$&9.84&1.65&646\\  
\hline  
C &2683&0.1280& 0.000009&0.2&$10^{-3}$&1.80&$59805$&14.36&11.40&796 \\ 
\hline  
D &2828&0.1281& 0.000007&0.2&$10^{-3}$&2.00&2301&17.16&28.60&896 \\
\hline
E &2966&0.1282& 0.000006&0.2&$10^{-3}$&2.20&0.37&94.73&278.63&996 \\
\hline    
F &3535&0.1285&$0.000003$&0.2&$10^{-3}$&3.14&0.004&198.07&750.85&1475 \\ 
\hline  
G &2449&0.1278&0.000013 &1.2&$10^{-3}$&1.50&$370801$&10.00&6.02&720 \\ 
\hline
\end{tabular*}  
\end{table}  
The above discussions did not offer a way to judge the period in which CSIPT happened, and here is our procedure. First we calculate $S(T)$ and next assume the RD era to determine $T_n$ via $S(T)\simeq 140$.  If indeed the ratio $\alpha_n\equiv \rho_0/\rho_r(T)$ at $T=T_n$ is smaller than 1, then the assumption is justified. Otherwise, CSIPT should complete in the vacuum dominance era and finally we take the criterion  $S(T)\simeq70$ to determine $T_n$. This is a simplified procedure, and we refer to Ref.~\cite{Ellis:2019oqb} for a more accurate treatment using iteration. In general their difference is not significant except for the subtle case where $\alpha_n$ is close to 1, and hence the era has comparable radiation and vacuum energy density. As a consequence, either criterion works well. We will go back to this point in a later concrete example.

Now we present the numerical results of CSIPT. We choose a few benchmarks points, which satisfy all the phenomenological constraints and requirements from radiative CSISB and DM discussed before; the condition Eq.~(\ref{cri:3}) is also imposed. Then only one free parameter $\ld_{sx}$ is left except for the irrelevant (to those phenomenologies) ones $\ld_x$ and $\ld_{hx}$. In Table.~\ref{Tn:GW} we show the benchmarks points in the GW scenario. 
One can see that $T_n$ increases with $\ld_{sx}$, which is traced back to the decreasing $S(T)$, explicit in Eq.~(\ref{S3:com}): In simple terms, the larger quantum (also thermal) correction benefits thermal tunneling. This behavior is explained by the narrower of the barrier, i.e., the shorter escaping path~\footnote{ Although not shown here, we find that  at the same time the barrier becomes shallower, which brings an opposite effect to $S(T)$, but it is supposed to be subdominant to the former effect.}, with the increasing $\ld_{sx}$; one can find its evidence in Fig.~\ref{GWvality}. Note that to make $T_n$ lie above the QCD chiral symmetry breaking scale, $\ld_{sx}$ should be sufficiently large, for instance $\ld_{sx}\gtrsim 1.44$ in the GW scenario.

The observed $T_n-\ld_{sx}$ behavior has immediate implications to CSIPT thus GW, and we can clearly see this from the table. Among the eight benchmarks, CSIPT of A, B, C, D and G, which have a relatively smaller $\ld_{sx}\lesssim 2.0$, completed in the vacuum dominated period, and they give a very large $\alpha_n$, characterizing strong supercooling. By contrast, CSIPT of E, F, which have a relatively larger $\ld_{sx}\gtrsim 2.2$,  completed in the RD era and give a suppressed $\alpha_n$. Therefore, the heavier DM region may be characterized by less vacuum energy release. This is not a good news since the heavier DM region is just the region which tends to go beyond the sensitivity of DM direct detection experiments.

We also display the benchmarks for the Higgs portal scenario in Table.~\ref{Higgsportal}, to find that the  two scenarios share fairly similar feature of CSIPT, provided that the values of $\ld_{sx}$ are close. It is not surprising since CSIPTs in both scenarios are dominated by the singlet scalar, whose quantum corrections dominantly come from the DM field. 
\begin{table}  
\caption{ Benchmark points in the Higgs portal scenario}  \label{Higgsportal}
\begin{tabular*}{14.2cm}{|p{0.5cm}|p{1.5cm}|p{1.1cm}|p{1.7cm}|p{1.5cm}|p{1cm}|p{1.3cm}|p{1.4cm}|p{1.4cm}|p{1.4cm}|}  
\hline  
 & $v_s/\rm GeV$&$\lambda$&$ \lambda_s$&$ \lambda_{hs}$&$\lambda_{sx}$&$\alpha_n$&$\wt\beta$&$T_n$/GeV&$T_{*}$/GeV\\  
\hline  
a &2245&0.1304&-0.00110&-0.00286&1.38&$3.0*10^8$&10.72&1.05&576\\ 
\hline  
b &2449&0.1299&-0.00119&-0.00262&1.50&$1.4*10^7$&14.75&2.43&688\\  
\hline  
c &2683&0.1294&-0.00142&-0.00218&1.80&$16213$&16.59&15.80&829\\ 
\hline  
d &2828&0.1293&-0.00158&-0.00196&2.00&0.78&82.49&28.60&923\\
\hline
e &2966&0.1292&-0.00174&-0.00178&2.20&0.30&99.90&293.83&1018\\
\hline    
f &3535&0.1291&-0.00249&-0.00124&3.14&0.003&203.46&753.56&1473\\ 
\hline  
\end{tabular*}  
\end{table}

\section{Abundant gravitational wave (GW) from CSIPT}

In the last section we have shown that CSIPT, due to the vanishing quadratic term of the scalon, is first order and moreover characterized by significant supercooling for the not very heavy trigger. So, the bubble collisions near the end of CSIPT stimulate abundant emission of GW, which may be hunted by eLISA, Tianqin, etc. From the DM direct detection bounds shown in Fig.~\ref{DD:Higgsportal} and Fig.~\ref{Fig_CW}, it is seen that the multi-TeV DM region is buried underneath the neutrino floor, and consequently it can not be probed by the DM direct detection experiments. Fortunately, the GW signal opens a window to probe this region. 


In estimating the GW spectra, there are two critical parameters which characterize first order PT, namely the $\alpha$ and $\wt \beta$ parameters defined as 
\begin{align} \label{} 
&\alpha\equiv \frac{\Delta\epsilon}{\rho_r}|_{T=T_{n}},\ \  \Delta\epsilon=\rho_0+T\frac{d}{dT}[V_{eff}(\phi_{0},T)-V_{eff}(0,T)],
\\
& \wt \beta\equiv -\f{1}{H}\f{dS(t)}{dt}|_{t=t_{n}}=T_{n} \frac{dS(T)}{dT}|_{T=T_{n}}.\label{betat}
\end{align}
$\alpha$ denotes the latent heat release $\Delta\epsilon$ normalized by the energy density of  radiation during PT. It receives two contributions, but in the strong supercooling PT it is obviously dominated by $\rho_0$, the vacuum energy difference defined in Eq.~(\ref{VG}). While $\wt\beta^{-1}\sim\tau_{PT}/ \tau_{H}$ denotes the time scale of PT duration, normalized by the Hubble time scale $\tau_H\sim 1/H$ at $t_n$. The GW amplitude is enhanced by the larger $\alpha$ and $\wt\beta^{-1}$. Their values have been listed in the previous tables.

One may obtain an overall picture about $\wt\beta^{-1}$. The typical behavior of $S(T)$ is plotted in Fig.~\ref{witten:app}, which leads us to the observation: At the relatively high temperature region $S(T)$ is almost linear in $T$, whose slope $\sim {\cal O}(0.5){\rm GeV}^{-1}$ merely slowly increases with decreasing $T$, however, the slope sharply increases when $T$ drops below certain temperature, and this trend becomes more significant with the lower $T$. Thereby, if CSIPT is completed at a higher $T_n$, one has $\wt \beta\sim {\cal O}(0.5) T_n/{\rm GeV}$; otherwise it may be enhanced by orders of magnitude. This observation roughly explains the pattern of $\wt \beta$ in Table.~\ref{Tn:GW} and~\ref{Higgsportal}.


%
%

%

%

\subsection{GW sources}

According to the present understanding of the GW emission during PT proceeding via the thermal bubble nucleation, there are three sources after bubble collision at $T_n$:
\begin{description}
\item[Bubble collision]  Before the bubble wall reaching the terminal velocity, almost all of the vacuum energy (or latent heat) will be transformed into the kinetic energy of the bubble wall. If the bubble wall is expanding in the vacuum, it runs away, that is to say, it keeps accelerating utile bubble collision. Even expanding in a plasma, the bubble was still believed to run away in the strongly supercooling PT with 
\begin{equation} \label{} 
\alpha>\alpha_{\infty}\equiv\f{\Delta P_{\rm LO}}{\rho_R}\approx \f{30}{24\pi^2}\f{\sum_a c_a \Delta m_a^2(\phi_n)}{g_*T_n^2}\sim 10^{-2}\L\f{\phi_n}{T_n}\R^2.
\end{equation}
Nevertheless, recently it is found that the friction on the wall at the next-to leading order is proportional to the Lorentz factor of the wall, $\Delta P_{\rm NLO}\propto \gamma$~\cite{NLO:P}. It is able to balance the wall when $\gamma\ra \gamma_{\rm eq}$, thus stopping runaway. Then, the energy stored in the bubble is still negligible provided that $\alpha$ does not become extremely large~\cite{Ellis:2019oqb}, far larger than the $\alpha$ considered in this paper. So, the GW source as usual is from the bulk motion of the plasma.


\item[Sound wave]  The first bulk motion is the sound wave propagating in the plasma after percolation happens. The fraction of latent heat that goes into the fluid motion is estimated to be~\cite{soundwave}
\begin{equation} \label{} 
\kappa_{sw}\approx \alpha(0.73+0.083\sqrt{\alpha}+\alpha)^{-1}\xrightarrow{\alpha\gg 1} 1.
\end{equation}
The GW peak frequency at $T_n$ is not well-understood, and is $f_{sw,*}=2/\sqrt{3}(8\pi)^{1/3}/R_*$ with $R_*$ the average bubble separation at collision. In the exponential approximation of $S(T)$, it is related to the typical time scale of PT: $R_*=\beta_n/v_w$. Redshifting to today, the observed peak is $f_{sw}=f_{sw,*}a_0/a(T_n)$ and parameterized as
\begin{equation} \label{} 
\begin{split}
f_{sw}=1.9\times10^{-5}\frac{\wt\beta}{v_{w}}\L\frac{T_n}{100 \rm GeV}\R\L\frac{g_*}{100}\R^{\frac{1}{6}} \rm Hz.
\end{split}
\end{equation}
The GW spectrum of the sound wave is 
\begin{align} \label{} 
h^2\Omega_{sw}(f)&=2.65\times 10^{-6}\f{1}{\wt\beta}\L\frac{ \kappa_{sw}\alpha}{1+\alpha}\R^{2}\L\frac{100}{g}\R^{\frac{1}{3}}v_{w} S_{sw}(f),\\
S_{sw}(f)&=(f/f_{sw})^3\left[\frac{7}{4+3f/f_{sw}}\right]^{\frac{7}{2}}.
\end{align}
For $\alpha\gg 1$ the GW enhancement by strong supercooling is saturated because the explicit $\alpha$ dependence in the spectrum is canceled. Then the GW spectrum is characterized by the single parameter $\wt \beta$.

\item[MHD turbulence] Percolation generates another fluid bulk motion, the MHD turbulence. It is supposed to have a suppressed efficiency factor $\kappa_{turb}\sim0.05\kappa_{sw}$~\cite{simulations} if the SW period could last over at least one Hubble time scale, namely $R_nU_f>1/H_n$ with the root-mean-square fluid velocity~\cite{simulations}
\begin{equation} \label{} 
U_f\simeq \f{\sqrt{3}}{2}\L\f{\alpha}{1+\alpha}\kappa_{sw}\R^{1/2}\xrightarrow{\alpha\gg 1} \f{\sqrt{3}}{2}.
\end{equation}
Otherwise $\kappa_{turb}$  may be significantly enhanced and becomes the dominant source. The GW spectrum of this source has peak frequency similar to that of the SW source, 
\begin{equation} \label{} 
\begin{split}
&f_{turb}=2.7\times10^{-5}\frac{\wt\beta}{v_{w}}\frac{T}{100 \rm GeV}\L\frac{g}{100}\R^{\frac{1}{6}}\rm Hz.
\end{split}
\end{equation}
And the GW spectrum is given by
\begin{equation} \label{} 
\begin{split}
&h^2\Omega_{turb}(f)=3.35\times10^{-4}\f{1}{\wt\beta}\L\frac{ \kappa_{turb}\alpha}{1+\alpha}\R^\frac{3}{2}\L\frac{100}{g}\R^{\frac{1}{3}}v_{w} S_{turb}(f),
\end{split}
\end{equation}
with the shape function
\begin{equation} \label{} 
\begin{split}
S_{turb}(f)=\frac{(f/f_{turb})^3}{[1+(f/f_{trub})]^{\frac{11}{3}}(1+8\pi f/h)}.
\end{split}
\end{equation}
which, compared to $S_{sw}(f)$, shows a moderately large suppression $\sim{\cal O}{(10)}$ in the high frequency region. 
\end{description}
We have to stress that all of these ``data-driven" expressions are reliable only for a weaker phase transition $\alpha\lesssim 0.1$. For very large $\alpha$ thus ultra relativistic bubbles, they are far beyond the ability of the current numerical simulation. Recently there are works towards analytical understanding of the GW in this limiting situation~\cite{Jinno:2019jhi}.


\subsection{Prospects of the GW signal }

\begin{figure}[htbp] 
\centering 
\includegraphics[height=6cm, width=12cm]{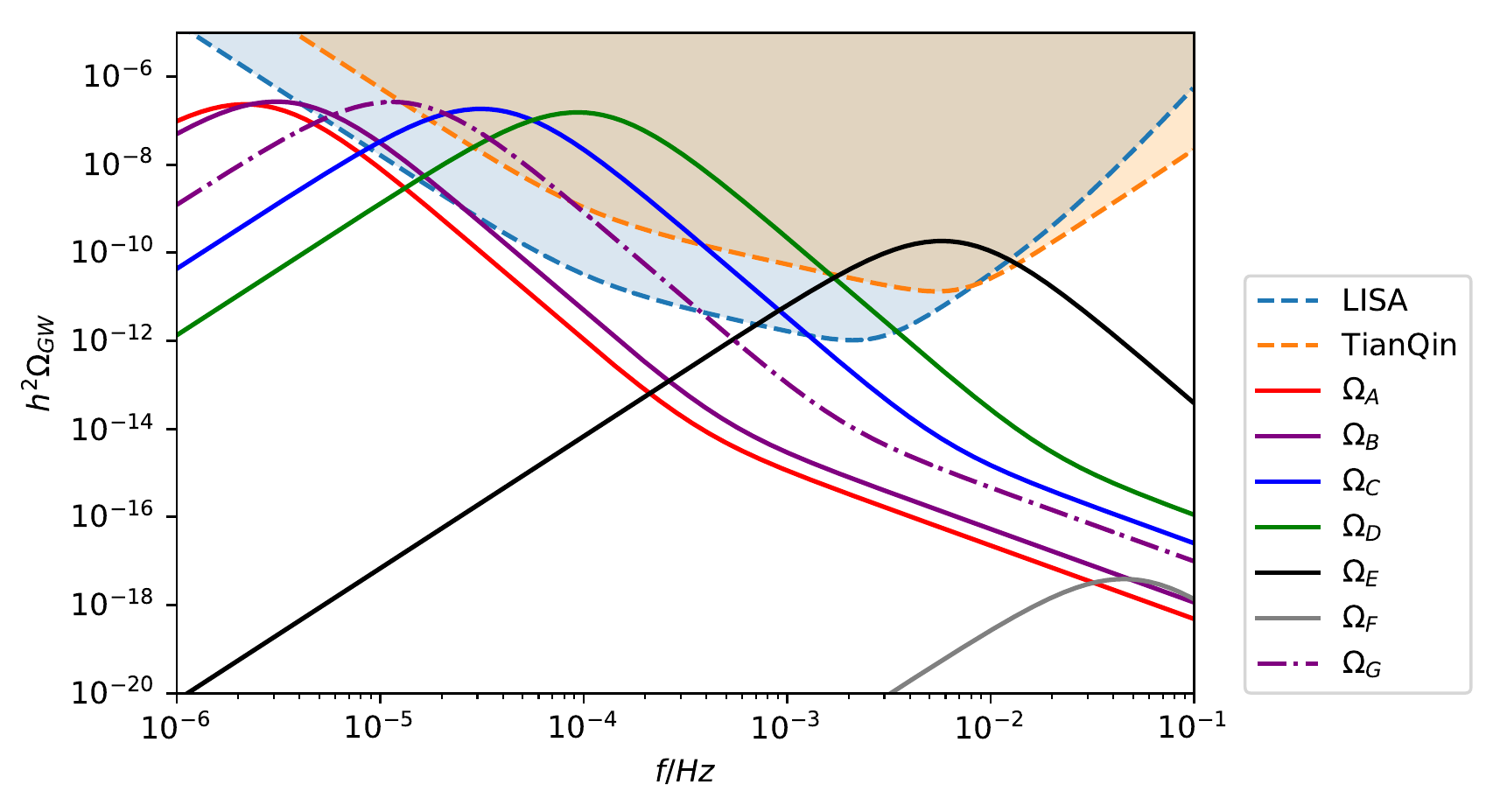} 
\caption{Detecting prospects of the (total) gravitational wave spectra of the benchmark points listed in Table~\ref{Tn:GW}, corresponding to the GW scenario.} \label{GW:HP}
\end{figure}

Now we have collected all the ingredients to demonstrate the tentative prospects of GW signatures of CSIPT by the DM. As an example, in Fig.~\ref{GW:HP}  we show the GW spectra of the benchmarks given in Table.~\ref{Tn:GW}, for the GW scenario. The sensitivity curves for TianQin and LISA~\cite{Moore:2014lga,Lu:2019sti} are plotted as the boundaries of the shaded regions. Only the spectra of two limiting benchmarks lie below the sensitivity  curves. One is F, which has a quite large $\ld_{sx}$ hence very effective bubble nucleation, weakening supercooling then giving a suppressed $\alpha\sim 10^{-2}$. The other one is A, which by contrast has a quite small $\ld_{sx}$, leading to a very low $T_n\sim 1$ GeV thus a low peak frequency $\sim 10^{-6}$Hz.

Additionally, it is of interest to notice that  the DM self interaction coupling $\lambda_x$, which basically is an irrelevant parameter in the zero temperature physics, can affect CSIPT through the daisy term, i.e., $\Pi_X\propto \ld_x T^2$; see Eq.~(\ref{self0}). As an illustration we set up B and G differing only in $\ld_{x}$. Increasing $\ld_x$, like increasing $\ld_{sx}$, helps to lift the bubble nucleation rate thus giving a higher $T_n$. Because $\alpha$ is already very large and its dependence in the spectra has been cancelled, then the spectra of B, whose $T_n$ is lower then a lower peak frequency, tends to move beyond the sensitivity region.  

%


%

\begin{figure}[htbp] 
\centering 
\includegraphics[height=6cm, width=12.1cm]{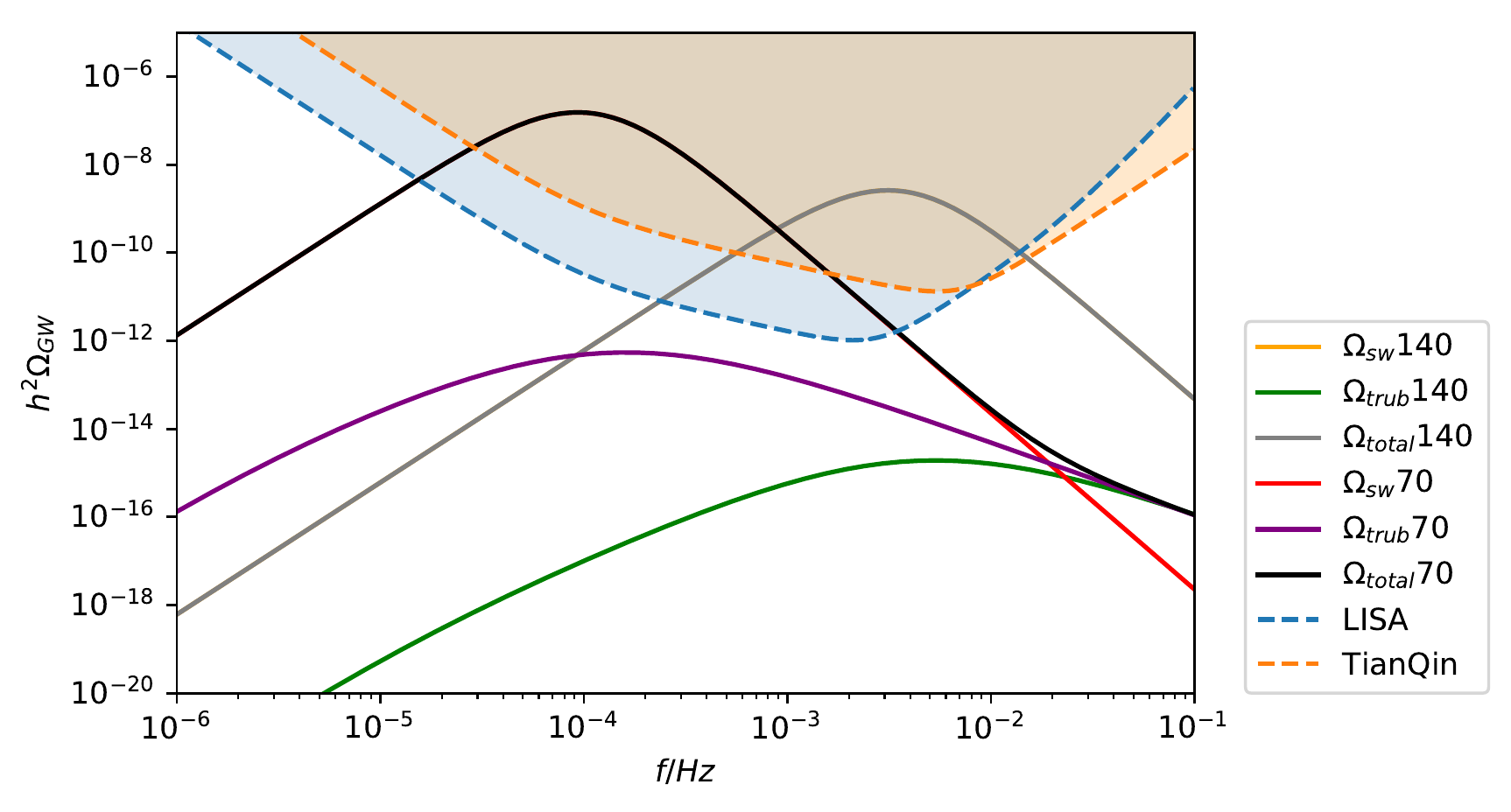} 
\caption{The gravitational wave spectra from Table~\ref{alpha1}, a subtle case to judge the CSIPT completion era, thus using both criteria for comparison. We display all three sources for gravitational wave.} \label{GW:GW}
\end{figure}

We end up with a comment on the subtle case $\alpha\sim 1$, which is ambiguous to determine the era when CSIPT completes. We show an example point using two different PT completion criteria in Table.~\ref{alpha1}. Following the crude rule in Section.~\ref{results}, we find that $\alpha$ is very close to 1 using the RD criterion $S(T)\simeq 140$. It indicates that the universe is transiting from the RD to the vacuum dominance era, so it should not be a very precise criterion. Then we also calculate CSIPT taking the vacuum dominance criterion, and one can see the sharp difference between the resulting PT parameters: $T_n$ jumps from 196.1 GeV to 28.6 GeV, and as a consequence the GW spectra significantly shifts to the IR frequency region, i.e., from the gray line to the black line in Fig.~\ref{GW:GW}. The actual CSIPT completion condition is neither $S(T)\simeq 140$ nor $S(T)\simeq 70$ but some value between them, and thus the actual GW spectrum should be located between the two spectra. To develop a more appropriate criteria for CSIPT completion is not trivial, and we leave it for a specific discussion elsewhere.

\begin{table}  
\caption{A subtle case of CSIPT completion criterion}  \label{alpha1}
\begin{tabular*}{16.35cm}{|p{2cm}|p{1.5cm}|p{1.1cm}|p{1.7cm}|p{1cm}|p{1cm}|p{1cm}|p{1.3cm}|p{1.4cm}|p{1.4cm}|p{1.4cm}|}  
\hline  
 & $v_s/\rm GeV$&$\lambda$&$ \lambda_s$&$ \lambda_x$&$\lambda_{hx}$&$\lambda_{sx}$&$\alpha$&$\wt\beta$&$T_n$/GeV&$T_{*}$/GeV\\  
\hline  
Radiation &2828&0.1281& 0.000007&0.2&$10^{-3}$&2.00&1.04&77.76&196.11&896 \\
\hline  
Vacuum &2828&0.1281& 0.000007&0.2&$10^{-3}$&2.00&2301&17.16&28.60&896 \\  
\hline  
\end{tabular*}  
\end{table}

\section{Conclusion and discussions}

The origin of the weak scale is a fundamental question in the SM. One attractive idea is imposing scale invariance  on the classical Lagrangian of some extension to SM, and then a scale is generated at the quantum level due to the anomaly of CSI. But the realistic CSI extension  to the SM needs a bosonic trigger, which is assigned to the scalar DM $X$ in this paper. This scenario establishes a direct connection between DM and scale genesis. To accommodate successful  DM phenomenologies, the radiative CSISB scale should $\sim {\cal O}(\rm TeV)$, which means that the GW from CSIPT, with the tendency of a large supercooling, can leave signals at the GW detectors such as LISA and Tianqin. Our analysis of GW signal is based on the usual appoarch. Recently, Ref.~\cite{Alanne:2019bsm,Schmitz:2020syl} presented a substaintially improved analysis, which is based on the noval peak-integrated sensitivity curves designed specifically for the GW from the strong first order PT. If applied, it may enhance detective prospect of our model.

Besides the overall physical picture, we pay great attentions to several aspects of techniques that are commonly used but not very clear, summarized in the following:
\begin{itemize} 
\item In the GW scenario, we in Section~\ref{GW:full} discuss if the strong quantum corrections could give rise to a significant difference between CSIPT analyze based on the single field along the tree level flat direction and full tunneling. Our numerical examples indicate that the tunneling path may be changed but the resulting difference in $S(T)$ is tolerable for normal couplings.
\item In Section~\ref{GW:full} we estimate the quality of Witten's formula which is frequently used  to calculate the nucleation rate in CSIPT, for the one-field case, and find that it is not a very good approximation, owing to the neglect of cubic term in high temperature expansion. We stress that the essence of Witten's approximation is the observation of validity of  high temperature expansion (to the quartic term) for CSIPT at very low temperature. Furthermore, we argue that it may also apply to the multi-field case.

\item In Section~\ref{criteria} we analyzed the completion condition for CSIPT with a very strong supercooling, which may make CSIPT completion happen in the vacuum dominated era rather than the ordinary RD era. We derive the analytical conditions for $S(T)$ in both cases, taking various criteria.

 \end{itemize} 
We are not content with these studies, for instance, a reliable condition for CSIPT in some subtle cases still requires further study.

\noindent {\bf{Acknowledgements}}

We would like to thank Yizhou Lu, Xuenan Chen and Xiangsong Chen for helpful discussions and specially thank Prof Xiangsong Chen for reading the first edition of this paper. This work is supported in part by the National Science Foundation of China (11775086).

\appendix

\section{Analytical solutions to RGEs}\label{RGESOL}

In this appendix we pursue an analytical approximation to Eq.~(\ref{RGE}), which consists of three
coupled RGEs. We impose the following hierarchy 
\begin{equation} \label{} 
\begin{split}
\lambda_{sx}\gg\lambda_x\gg\lambda_s.
\end{split}
\end{equation}
This hierarchy is reasonable. First, $\lambda_x\gg\lambda_s$ explains why radiative correction drives the $S$ rather than $X$ away from the origin. That is to say, this hierarchy guarantees the stability of DM field $X$. Next, $\lambda_{sx}\gg\lambda_s$ is the usual condition for raidative symmetry breaking. Third, 
since $\lambda_{sx}$ is already relatively large, a smaller $\ld_x$ is good for perturbativity~\footnote{In principle we have no compelling arguments to exclude the opposite pattern $\ld_{x}\gg \ld_{sx}$. Actually it is of interest to explore if radiative symmetry breaking can be driven by a large DM self-interaction $\ld_x$, basically a two loop effect on $\ld_s$. Such a scenario has not been discussed yet.}. With the above hierarchy, RGEs in Eq.~(\ref{RGE}) are reduced to a situation similar to scalar QED at the leading order of $\ld_{x}/\ld_{sx}$. Then the solution takes the form of~\cite{Coleman:1973jx}
\begin{equation} \label{SRGE} 
\begin{split}
&\lambda_{sx}(t)=\f{\lambda_{sx}(0)}{1-\f{\lambda_{sx}(0)}{8\pi^2}t},
\\
&\lambda_{s}(t)=\f{b}{2a}\lambda_{sx}(t)+\f{X}{a}\lambda_{sx}(t)\tan\left(\f{X}{b}\log\f{\lambda_{sx}(t)}{\pi}+A\R,
\end{split}
\end{equation}\label{}
where $t=\log \f{\mu}{\mu_0}$ with $\mu_0$ is the renormalization scale; $A=\arctan\f{a[\lambda_s(0)-\f{b}{2a}\lambda_{sx}(0)]}{X\lambda_{sx}(0)}-\f{X}{b}\log\f{\lambda_{sx}(0)}{\pi}$ with $X=\sqrt{ac-\f{b^2}{4}}$ in which the constants $a=\f{27}{2\pi}$, $b=\f{1}{4\pi}$, $c=\f{9}{8\pi}$. 

\section{The failure of common high temperature expansion}\label{AA}

In this appendix we give an example to show that in analyzing CSIPT the usual high temperature expansion keeping only the quadratic term~\cite{Marzola:2017jzl,Ghorbani:2017lyk} may leads to a sizable error in calculating $S(T)$. We consider the GW scenario of our model, and the effective potential in the high temperature expansion to the quadratic term is given by
\begin{equation} \label{FT:GW} 
\begin{split}
&\ \ \ \ \ \ \ \ \ \ \ \ \ \ \ \ \ \ V_{HT}(\phi,T)=V_0(\phi)+V_0^{(1)}(\phi)+C T^2 \phi^2,\\
&C=\frac{1}{12 \left\langle\phi\right\rangle} \left(m_{\phi_1}^2(\vec n)+m_X^2(\vec n)+6m_W^2(\vec n)+3m_Z^2(\vec n)+6m_t^2(\vec n)\right).
\end{split}
\end{equation}
Our numerical example is the point E in Table.~\ref{Tn:GW}. Using $V_{TH}$, we obtain that the CSIPT completion temperature is $18\rm GeV$, in the vacuum dominated era. By contrast, the CSIPT is found to be completed in the RD era, at 278 GeV, if we use the complete effective potential. Therefore, the wrongly used high temperature may leads to dramatic difference in CSIPT.


%
%
%
%
%
%

\end{document}